%% file: gampen.tex
\documentclass[twocolumn]{aastex63}

\graphicspath{{./}{figures/}} %%to tell LateX to also look for figures in the figures folder

\newcommand\sersic{S\'ersic}

\newcommand\gamornet{G\textsc{a}M\textsc{or}N\textsc{et}}
\newcommand\gampen{GaMPEN}

\shorttitle{Galaxy Morphology Posterior Estimation Network: An Overview}
\shortauthors{Ghosh et al.}

\usepackage[normalem]{ulem}
\usepackage{enumitem}
\usepackage{bm}
\usepackage{amsmath}
\usepackage{gensymb}
\usepackage{subfigure}
\usepackage{multirow}
\usepackage{array}
\usepackage{hhline}
\usepackage{hyperref}
\PassOptionsToPackage{hyphens}{url}\usepackage{hyperref}

\usepackage{CJK}
\input{bangla_commands}

\begin{document}
\begin{CJK*}{UTF8}{gbsn}

\title{GaMPEN: A Machine Learning Framework for Estimating Bayesian Posteriors of Galaxy Morphological Parameters}

\author[0000-0002-2525-9647]{Aritra Ghosh ({\bngxi Airt/r \*gh*eaSh})}
\affil{Department of Astronomy, Yale University, New Haven, CT, USA}
\affil{Yale Center for Astronomy and Astrophysics, New Haven, CT, USA}
\email{aritra.ghosh@yale.edu; aritraghsh09@gmail.com}

\author[0000-0002-0745-9792]{C. Megan Urry}
\affil{Yale Center for Astronomy and Astrophysics, New Haven, CT, USA}
\affil{Department of Physics, Yale University, New Haven, CT, USA}

\author{Amrit Rau}
\affil{Department of Computer Science, Yale University, New Haven, CT, USA}

\author[0000-0003-3544-3939]{Laurence Perreault-Levasseur}
\affil{Department of Physics, Univesité de Montréal, Montréal, Canada}
\affil{Mila - Quebec Artificial Intelligence Institute, Montréal, Canada}
\affil{Center for Computational Astrophysics, Flatiron Institute, New York, NY, USA}

\author[0000-0002-6458-3423]{Miles Cranmer}
\affil{Department of Astrophysical Sciences,
Princeton University, Princeton, NJ, USA}

\author{Kevin Schawinski}
\affiliation{Modulos AG, Technoparkstr. 1, CH-8005, Zurich, Switzerland}

\author{Dominic Stark}
\affiliation{Modulos AG, Technoparkstr. 1, CH-8005, Zurich, Switzerland}

\author[0000-0003-4056-7071]{Chuan Tian (田川)}
\affil{Yale Center for Astronomy and Astrophysics, New Haven, CT, USA}
\affil{Department of Physics, Yale University, New Haven, CT, USA}

\author{Ryan Ofman}
\affil{Department of Astronomy, Yale University, New Haven, CT, USA}

\author[0000-0001-8211-3807]{Tonima Tasnim Ananna}
\affiliation{Department of Physics and Astronomy, Dartmouth College, 6127 Wilder Laboratory, Hanover, NH, USA}

\author[0000-0002-5504-8752]{Connor Auge}
\affiliation{
Institute for Astronomy, University of Hawai`i, Honolulu, HI, USA}

\author[0000-0002-1697-186X]{Nico Cappelluti}
\affiliation{Department of Physics, University of Miami, Coral Gables, FL, USA}
\affiliation{INAF - Osservatorio di Astrofisica e Scienza dello Spazio di Bologna, Bologna, Italy}

\author[0000-0002-1233-9998]{David B. Sanders}
\affiliation{
Institute for Astronomy, University of Hawai`i, Honolulu, HI, USA}

\author[0000-0001-7568-6412]{Ezequiel Treister}
\affiliation{Instituto de Astrofísica, Facultad de Física, Pontificia Universidad Católica de Chile, Santiago, Chile}

\begin{abstract}
We introduce a novel machine learning framework for estimating the Bayesian posteriors of morphological parameters for arbitrarily large numbers of galaxies.
The Galaxy Morphology Posterior Estimation Network (\gampen{}) 
estimates values and uncertainties for a galaxy's bulge-to-total light ratio ($L_B/L_T$), effective radius ($R_e$), and flux ($F$). To estimate posteriors, \gampen{} uses the Monte Carlo Dropout technique and incorporates the full covariance matrix between the output parameters in its loss function. %The latter accounts for covariances among the output parameters, and thus the predicted posteriors can be calibrated properly.  
\gampen{} also uses a Spatial Transformer Network (STN) to automatically crop input galaxy frames to an optimal size before determining their morphology. This will allow it to be applied to new data without prior knowledge of galaxy size.
Training and testing \gampen{} on galaxies simulated to match $z < 0.25$ galaxies in Hyper Suprime-Cam Wide \textit{g}-band images, we demonstrate that \gampen{} achieves typical errors of $0.1$ in $L_B/L_T$, $0.17$ arcsec ($\sim 7\%$) in $R_e$, and $6.3\times10^4$ nJy ($\sim 1\%$) in $F$. \gampen{}'s predicted uncertainties
are well-calibrated and accurate ($<5\%$ deviation) -- for regions of the parameter space with high residuals, \gampen{} correctly predicts correspondingly large uncertainties.
%Testing shows that \gampen{} predictions become less precise for especially small ($R_e < 1\arcsec$) and/or faint ($F < 10^6$ nJy) galaxies,where \gampen{} correctly predicts correspondingly larger uncertainties.
We also demonstrate that we can apply categorical labels (i.e., classifications such as ``highly bulge-dominated") to predictions in regions with high residuals and verify that those labels are  $\gtrsim 97\%$ accurate. To the best of our knowledge, \gampen{} is the first machine learning framework for determining joint posterior distributions of multiple morphological parameters and is also the first application of an STN to optical imaging in astronomy.

\end{abstract}


\vspace{10pt}

\section{Introduction} \label{sec:intro}
For almost a century, starting with \citeauthor{hubble_1926} in \citeyear{hubble_1926}, astronomers have linked the morphology of galaxies to the physics of  galaxy formation and evolution. Morphology has been shown to be related to many fundamental properties of the galaxy and its environment, including galaxy mass, star formation rate, stellar kinematics, merger history, cosmic environment, the influence of supermassive black holes \citep[e.g.,][]{Bender1992DynamicallyProperties,Tremaine2002TheCorrelation,pozzetti_10, wuyts_11, Schawinski2014TheGalaxies, Huertas-Company2016MassCANDELS,powell_17, Shimakawa2021Subaru0.30.6, Dimauro2022CoincidenceGrowth}. Studying the morphology of large samples of galaxies at different redshifts is crucial in order to understand the physics of galaxy formation and evolution. 

Over the last decade, machine learning (ML) has been increasingly employed by astronomers for a wide variety of tasks -- from identifying exoplanets to studying black holes \citep[e.g.,][]{ml_pz,ml_sz,Shallue2018IdentifyingKepler-90,Sharma2020ApplicationClassification,Natarajan2021QuasarNet:Holes}. Especially, Convolutional Neural Networks (CNNs)\footnote{CNNs are a specific form of machine learning algorithm that specializes in processing data with a grid-like topology, such as an image. See \S \ref{sec:cnn_description} for more details.} have revolutionized the field of image processing and have become increasingly popular for determining galaxy morphology \citep[e.g.,][]{Dieleman2015Rotation-invariantPrediction, Huertas-Company2015ALEARNING, Tuccillo2018DeepFitting, Hausen2020MorpheusData, Walmsley2020GalaxyLearning, Cheng2021GalaxyNetworks, Vega-Ferrero2021PushingSurvey, Tarsitano2022ImageLearning}. Previously, we have developed a publicly available CNN, called \gamornet{} \citep{Ghosh2020GalaxyGalaxies}, that classifies galaxies morphologically with minimal real training data, and has been demonstrated to achieve accuracy $\gtrsim 95\%$ across multiple datasets. 

This use of CNNs has been driven by the fact that traditional methods of classifying morphologies---visual classification and template fitting to the surface brightness profile of a galaxy---are not scalable to the data volume expected from future surveys such as The Vera Rubin Observatory Legacy Survey of Space and Time \citep[LSST;][]{lsst}, the Nancy Grace Roman Space Telescope \citep[NGRST;][]{ngrst}, and Euclid \citep{euclid}. The quality of fits obtained using template fitting depends significantly on the initial input parameters, and when dealing with millions of galaxies, such hand-refinement of input parameters is an intractable task. Although large citizen science projects like Galaxy Zoo \citep{gzoo_original} have been successful in processing many surveys in the past, even these will fail to keep up with the upcoming data glut. Moreover, reliable visual classifications require a decent signal-to-noise ratio, take time to set up and execute, and require an extremely careful de-biasing of the vote shares obtained \citep[e.g.,][]{gzoo_original,gzoo_candels}. 

From early attempts at using a CNN to classify galaxies morphologically (e.g.,  \citealp{Dieleman2015Rotation-invariantPrediction}) to the largest CNN produced morphology catalogs currently available \citep{Cheng2021GalaxyNetworks, Vega-Ferrero2021PushingSurvey}, most
CNNs have provided broad, qualitative classifications, rather than 
numerical estimates of morphological parameters. Such studies typically entail classifying galaxies based on their morphological properties (e.g., based on whether the galaxy has a disk or a bulge or a bar, etc.) as opposed to predicting values of relevant morphological parameters that help characterize the galaxy (such as bulge-to-total light ratio, radius, etc.).
By contrast, \citet{Tuccillo2018DeepFitting} used a CNN to estimate the parameters of a single-component \sersic{} fit, though  without uncertainties. 
Meanwhile, the computation of full Bayesian posteriors for different morphological parameters is crucial for drawing scientific inferences that account for uncertainty and thus are indispensable in 
the derivation of robust scaling relations  \citep[e.g.,][]{Bernardi2013TheProfile, vanderWel20143D-HST+CANDELS:3} or tests of theoretical models using morphology \citep[e.g.,][]{Schawinski2014TheGalaxies}. Thus, producing posterior estimates will significantly increase the scientific potential of morphological catalogs produced using CNNs.

In this work, we introduce \gampen{} (the Galaxy Morphology Posterior Estimation Network), a novel machine learning framework that estimates the Bayesian posteriors for three morphological parameters: the bulge-to-total light ratio ($L_B/L_T$), the effective radius ($R_e$), and the total flux ($F$).
\gampen{} uses a %downstream 
CNN module to estimate the joint posterior probability distributions of these parameters. This is done by using the negative log-likelihood of the output parameters as the loss function combined with the Monte Carlo Dropout technique \citep{gal_2016}. We also used the full covariance matrix in the loss function, using a series of algebraic manipulations (see \S \ref{sec:uncertainties}). The full covariance matrix accounts for dependencies among different output parameters and ensures that the posterior distributions for all three output variables are well calibrated.

Although the use of CNNs in the recent past has allowed astronomers to process large data volumes quickly, some challenges related to data pre-processing have remained. One of these challenges has to do with making cutouts of proper sizes. Most trained CNNs require input images of a fixed size---thus, most previous work (e.g., \citealp{Cheng2021GalaxyNetworks, Vega-Ferrero2021PushingSurvey}) has resorted to selecting a large cutout size for which ``most galaxies" would remain in the frame. However, this means that for many galaxies in the dataset, especially smaller ones, typical cutouts contain other galaxies in the frame, often leading to less accurate results. This problem is aggravated when designing a CNN applicable over an extensive range in redshift, which corresponds to a large range of galaxy sizes. Lastly, most previous work has used computations of $R_e$ from previous catalogs to estimate the correct cutout size to choose. This is, of course, not possible when one is trying to use a CNN on a new, unlabeled dataset. 

To address these challenges, \gampen{} automatically crops the input image frames using a Spatial Transformer Network (STN) module upstream of the CNN. STNs are self-consistent modules that can be used for the spatial manipulation of data within machine learning frameworks. In \gampen{}, based on the input image, the STN predicts the parameters of an affine transformation which is then applied to the input image. The transformed image is then passed onto the downstream CNN.
The inclusion of the STN in the framework greatly reduces the amount of time spent on data pre-processing as it trains simultaneously with the downstream CNN without additional supervision. We later show in \S \ref{subsec:STN} how the STN automatically learns to make appropriate affine transformations (such as cropping) on the input data, which are helpful in the downstream task of morphological parameter estimation.

To the best of our knowledge, \gampen{} is the first ML framework to apply an STN to optical imaging data and is the first to estimate full Bayesian posteriors for galaxy morphological parameters. In order to have a robust understanding of the performance, bias, and limitations of \gampen{}, we train and test \gampen{} on simulations of galaxy images---the only situation where we have access to the ``ground truth" morphological parameters of the galaxies. We match our simulations to the observations of the Hyper Suprime-Cam (HSC) Wide survey \citep{hsc_pdr1}, as this is an obvious application (to be described in a forthcoming paper).
%In a future paper, we will describe the application of \gampen{} to real HSC data. 
We use real HSC Wide images, with their multiples galaxies, to validate the STN performance.

In \S \ref{sec:sim_data}, we describe the simulated data used to train and test \gampen{}. We describe the structure and code of \gampen{} in \S \ref{sec:cnn_description} and outline the entire mechanism behind the prediction of posteriors in \S \ref{sec:uncertainties}. In \S \ref{sec:training} we describe \gampen{}'s training procedure. We present our results in \S \ref{sec:results}, and summarize our findings along with future applications of \gampen{} in \S \ref{sec:conclusions}. \gampen{}'s data-access policies are described in Appendix  \ref{sec:ap:data_access}. 

\section{Simulated Galaxies} \label{sec:sim_data}
We train and test \gampen{} using mock galaxy image cutouts simulated to match \textit{g}-band data from the Hyper Suprime-Cam (HSC) Subaru Strategic Program wide-field optical survey \citep{hsc_pdr1}. The Subaru Strategic Program, ongoing since 2014, uses the HSC prime-focus camera, which provides extremely high sensitivity and resolving power due to the large 8.2 meter mirror of the Subaru Telescope. Its \textit{g}-band seeing, with median FWHM of $0.85''$, is a large improvement over the Sloan Digital Sky Survey (SDSS; \citealp{sdss_tech_summary}), which has a median \textit{g}-band seeing of $1.4''$.

To generate mock images, we used GalSim \citep{Rowe2015GalSim:Toolkit}, the modular galaxy image simulation toolkit. GalSim has been extensively tested and shown to yield very accurate rendered images of galaxies. We simulated 150,000 galaxies in total, with a mixture of both single and double components, in order to have a diverse training sample. To be exact, $75\%$ of the simulated galaxies consisted of both bulge and disk components, while the remaining $25\%$ had either a single disk or a bulge. 

For both the bulge and disk components, we used the \sersic{} profile, the surface brightness of which is given by  

\begin{equation}
\Sigma(R) = \Sigma_e \exp \left[ -\kappa \left( \left( \frac{R}{R_e}\right)^{1/n} - 1 \right) \right] ,
\label{eq:sersic_fn}
\end{equation}

\noindent
where $\Sigma_e$ is the pixel surface brightness at the effective radius $R_e$, $n$ is the Se\'rsic index, which controls the concentration of the light profile, and $\kappa$ is a parameter coupled to $n$ that ensures that half of the total flux is enclosed within $R_e$.
The standard formula for an exponential disk corresponds to $n=1$, and a de\,Vaucouleurs profiles is $n=4$.

\begin{deluxetable*}{cccccc}[htbp]
%\tablenum{2}
\tablecaption{Parameter Ranges of Simulated Galaxies  \label{tab:sim_para}}
\tablecolumns{6}
\tablehead{
\colhead{Component Name} & \colhead{\sersic{} Index} & \colhead{Half-Light Radius} & \colhead{Flux} & \colhead{Axis Ratio} & \colhead{Position Angle} \\ 
\colhead{} & \colhead{} & \colhead{(arcsec)} & \colhead{(nJy)} & \colhead{} & \colhead{(degrees)}
}
\startdata
    \hline
    \hline
    \multicolumn{6}{c}{Single Component Galaxies} \\
    \hline
     & 0.8 - 1.2 or 3.5 - 5.0\tablenotemark{a} & 0.1 - 5.0 & $10^3$ - $5\times10^6$ & 0.25 - 1.0 & $-90.0$ - $90.0$ \\
    \hline
    \hline
    \multicolumn{6}{c}{Double Component Galaxies} \\
    \hline
    Disk & 0.8 - 1.2 & 0.1 - 5.0 & 0.0 - 1.0\tablenotemark{b} & 0.25 - 1.0 & $-90.0$ - 90.0\\
    %\hline
    Bulge & 3.5 - 5.0 & 0.1 - 3.0 & 1.0 - Disk. Comp.\tablenotemark{b} & 0.25 - 1.0 &  Disk Comp. $\pm\,[0,15]$\tablenotemark{c} \\
\enddata
\tablenotetext{a}{The single component galaxies are equally divided between galaxies with a \sersic{} index between 0.8 - 1.2 and galaxies with a \sersic{} index between 3.5 - 5.0.}
\tablenotetext{b}{Fractional fluxes are noted here. The bulge flux fraction is chosen such that for each simulated galaxy it is added with the disk flux fraction to give $1.0$. The total flux of the galaxies is varied between $10^3$ and $5\times10^6$ nJy.}
\tablenotetext{c}{The bulge position angle differs from the disk position angle by a randomly chosen value between $-15$ and $+15$ degrees.}
\tablecomments{The above table shows the ranges of the various \sersic{} profile parameters used to simulate training and testing data. $75\%$ of the simulated galaxies have both disk and bulge components, and the remainder have either a disk or a bulge component. The distributions of all the simulation parameters are uniform except for the position angle and flux of the double-component galaxies. For more details about these choices, please refer to  \S\,\ref{sec:sim_data}.}
\end{deluxetable*}

The parameters required to generate the \sersic{} profiles were drawn from uniform distributions, over ranges given in Table \ref{tab:sim_para}. For the disk and bulge components, we let the \sersic{} index vary between $0.8-1.2$ and $3.5-5.0$, respectively. We chose to have varying \sersic{} indices as opposed to fixed values for each component in order to have a training set with diverse light profiles. The parameter ranges for fluxes and half-light radii are quite expansive, such that the simulations represent most local galaxies \citep{binney_and_merrifield} at $z \leq 0.25$ (i.e., the simulation parameters are chosen to match HSC $z < 0.25$ galaxies).

Specifically, single-component galaxies were assigned a half-light radius between $0.25$ kpc and $11.5$ kpc. In the double-component galaxies, the disk half-light radius was varied across the same range, and the bulge half-light radius was varied between $0.25$ kpc and $7.0$ kpc. To obtain the corresponding angular sizes for simulation, we placed the sample at $ z = 0.125$ using the Planck18 cosmology ($H_0=67.7$ km/s/Mpc, \citealp{planck18}) and the appropriate pixel scale.

For the single-component galaxies, the total flux was varied between $10^3$ and $5\times10^6$ nJy ($m_{\mathrm{AB}}\sim14 - 23$). For the double component galaxies, we first draw $L_B/L_T$ from a uniform distribution between 0 and 1. Thereafter, the total flux of the galaxy is chosen from a uniform distribution with a range of $10^3 - 5\times10^6$ nJy. To assign fluxes to the bulge and disk components, we multiply $L_B/L_T$ and $(1 - L_B/L_T)$ respectively by the total flux. Not following this procedure and drawing the bulge and disk fluxes independently causes most galaxies in the training set to have a very high or a very low $L_B/L_T$, which is not the case for most galaxies, and in any case we already have single-component galaxies in our sample. For the double component sample, we wanted to have a sufficient number of galaxies with intermediate values of $L_B/L_T$.

What matters in training a CNN is not matching the observed distributions of the simulation parameters; rather, it is spanning the full range of those parameters. 
Having too many of any one type---even if that is the reality in real data---can result in lower accuracy for minority populations (e.g., \citealp{Ghosh2020GalaxyGalaxies}).
By not weighting the simulated galaxy sample in any specific regions of the parameter space, we are able to optimize \gampen{} for the full range of galaxy morophologies.
%a diverse set of galaxies and are also able to thoroughly investigate whether \gampen{} performs poorly in any specific regions of the parameter space.

\begin{figure}[htb]
    \centering
    \includegraphics[width = 0.47\textwidth]{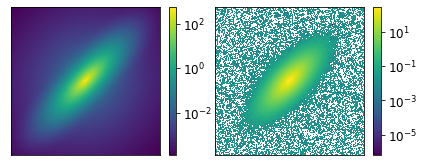}
    \caption{Two stages of simulating an HSC galaxy. (\textit{Left}): A randomly chosen two-dimensional light profile generated by GalSim. (\textit{Right}): The same image after PSF convolution and noise addition. The white pixels represent (small) negative values that arise from the process of noise addition.}
    \label{fig:sim_process}
\end{figure}

To make the two-dimensional light profiles generated by GalSim realistic, we convolved these with a representative point-spread function (PSF), and added appropriate noise. Figure \ref{fig:sim_process} shows a randomly chosen simulated light profile, as well as the corresponding image cutout generated after PSF convolution and noise addition. 

To curate a collection of representative PSFs, we first selected 100 galaxies at random from the HSC PDR2 Wide field \citep{hsc_pdr3} with $z \leq 0.25$ and $m_g \leq 23$, and that did not have any quality flags set to True (the quality flags check for cosmic ray hits, interpolated pixels etc.). We then used the HSC PSF Picker Tool\footnote{\href{https://hsc-release.mtk.nao.ac.jp/psf/pdr3/}{https://hsc-release.mtk.nao.ac.jp/psf/pdr3/}} to obtain the PSF at the location of these 100 galaxies. Each simulated light profile was convolved with a randomly chosen PSF out of these 100. To make sure that the PSFs are representative (i.e., do not contain any outliers), we ran a test where we convolved each one with a simulated galaxy light profile, before adding noise. 
We then inspected all possible difference images for each convolved galaxy, to make sure the average pixel value of the difference image was always at least three orders of magnitude lower than the average pixel value of the convolved galaxy image.

To generate representative noise, we used one-thousand $2\times2$ arcsec ``sky objects" from the HSC PDR2 Wide field. Sky-objects are empty regions identified by the HSC pipeline that are outside object footprints and are recommended for being used in blank-sky measurements. We visually verified that our sky objects did not contain any sources. We then read in the pixel values of these sky objects to generate a large sample of noise pixels. We randomly sampled this collection of noise pixels to make two-dimensional arrays of the same size as that of the simulated images and then added them to the images.

\begin{figure*}[htb]
    \centering
    \includegraphics[width
    =\textwidth]{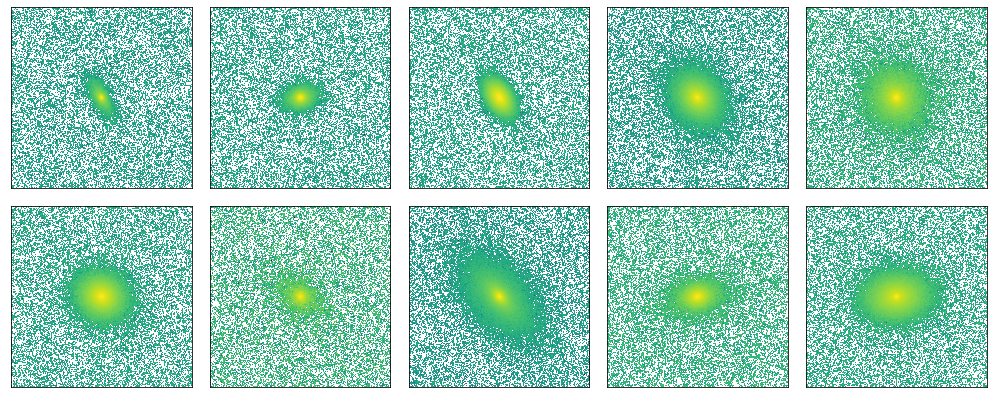}
    \caption{Ten randomly selected galaxies from our simulated dataset. The simulation parameters are chosen such that the simulated galaxies represent a diverse range of light profiles and include most bright, local galaxies at $z \lesssim 0.25$.}
    \label{fig:random_sim_galaxies}
\end{figure*}

All the simulated galaxy cutouts were chosen to have a size of $239 \times 239$ pixels, which translates to roughly $40 \times 40$ arcsecs given HSC's pixel scale of $0.168$ arcsecs/pixel. Ten randomly chosen simulated galaxies from our dataset are shown in Figure \ref{fig:random_sim_galaxies}.

\section{\gampen{} Architecture} \label{sec:cnn_description}

Artificial neural networks, consisting of many connected individual units called artificial neurons, have been studied for more than five decades. These artificial neurons are usually arranged in multiple layers and such networks typically have (a) an input layer to feed data into the network; (b) an output layer that contains the result of propagating the data through the network. In between, there are additional hidden layer(s). Each neuron is characterized by a weight vector $\bm{w}=(w_1,w_2,\ldots,w_n)$ and a bias $b$. The output of a single neuron in the network is given by 

\begin{equation}
y = \sigma(\bm{w} \cdot \bm{x} + b) ,
\label{eq:neuron_output}
\end{equation}

\noindent
where $\sigma$ is the chosen activation function of the neuron and $\bm{x}$ is the vector of inputs to the neuron. 
The process of training an artificial neural network involves finding the optimum set of weights and biases of all neurons such that, for a given set of inputs, the output of the network resembles the desired outputs as closely as possible. The optimization is usually performed by minimizing a loss function using stochastic gradient descent.

%\begin{figure}[htb]
%    \centering
%    \includegraphics[width
%    =0.4\textwidth]{schematic_toy_network.pdf}
%    \caption{A schematic diagram showing a simple %artificial neural network with a single hidden %layer. Image reproduced from %\cite{Ghosh2020GalaxyGalaxies}.}
%    \label{fig:toy_network_schematic}
%\end{figure}

The %form of artificial neural network that forms the 
backbone of \gampen{} is a Convolutional Neural Network (\citealp{fukushima_80, lecun_98}). %While traditional artificial neural networks learn global patterns in their input feature space, CNNs learn to identify thousands of local patterns in their input images that are translation invariant. 
Without convolutional layers, neural networks learn global patterns, whereas CNNs learn to identify thousands of local patterns in their input images that are translation invariant.
Additionally, CNNs learn the spatial hierarchies of these patterns, allowing them to process increasingly complex and abstract visual concepts. These two key features have allowed deep CNNs to revolutionize the field of image processing in the last decade \citep{lecun_15, dl_overview}. 

\begin{figure*}[htb]
    \centering
    \includegraphics[width
    =\textwidth]{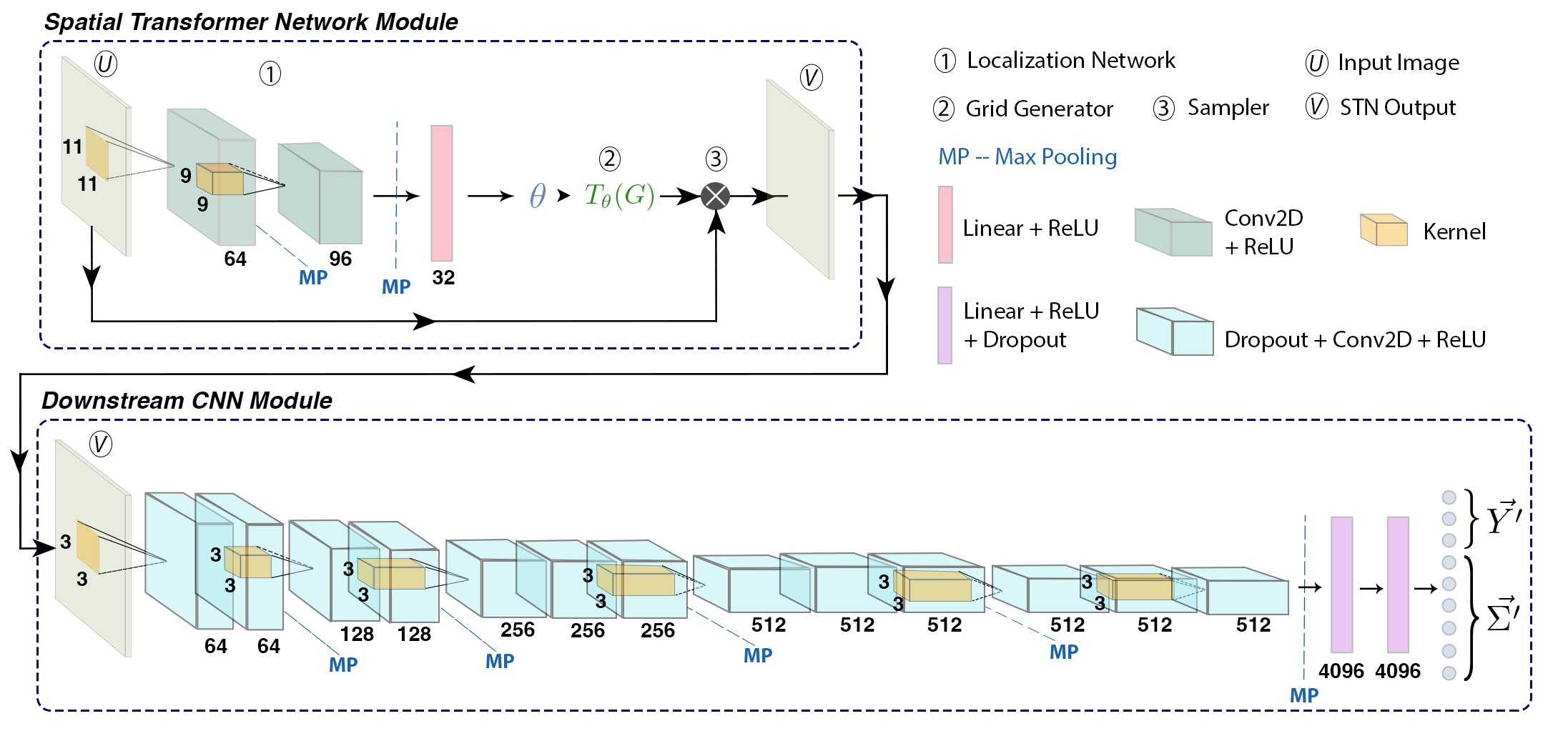}
    \caption{A schematic diagram of the Galaxy Morphology Posterior Estimation Network. \gampen's architecture consists of a downstream CNN module preceded by an upstream STN module. The CNN module empowers \gampen{} to estimate posterior distributions of galaxy morphology parameters. The upstream STN module trains without any extra supervision and learns to apply appropriate cropping transformations to the input image before passing it on to the CNN (for more details about these modules, see \S\S\,\ref{subsec:STN}, \ref{subsec:cnn}).
    The numbers below each layer refer to the number of filters/neurons in each layer. The yellow boxes inside the convolutional layers show the kernel and the number beside it refers to the corresponding kernel size. Only one kernel is shown per set of convolutional layers; all other layers in the set have kernels of the same size. Conv2D and ReLU refer to Convolutional Layers and Rectified Linear Units, respectively (described in \S.\ref{subsec:cnn}).}
    \label{fig:gampen_schematic}
\end{figure*}

The architecture of \gampen{} is shown in Figure\,\ref{fig:gampen_schematic}. It consists of a Spatial Transformer Network module followed by a downstream CNN module, described in \S\,\ref{subsec:STN} and \ref{subsec:cnn}, respectively. The design of \gampen{} is based on our previously successful classification CNN, \gamornet{} \citep{Ghosh2020GalaxyGalaxies}, as well as different variants of the Visual Geometry Group (VGG) networks \citep{vgg}, which are highly effective at large-scale visual recognition. We tried different architectures of these ``base" models by varying the depth of the entire network and the sizes of the various layers. To quickly and systematically search this model-design space, we use ModulosAI's\footnote{\href{https://www.modulos.ai}{https://www.modulos.ai}} AutoML platform, which uses a Bayesian optimization strategy. The said strategy involves using the current model's performance to determine which variant to try next. When choosing new configurations, the optimizer balances the exploitation of well-performing search spaces and the exploration of unknown regions.

To implement \gampen{}, we use PyTorch,
%\footnote{\href{https://pytorch.org}{https://pytorch.org}}
 which is an open-source machine learning framework, written in Python.

\subsection{The Spatial Transformer Network Module } \label{subsec:STN}
Spatial Transformer Networks (STNs) were introduced by \cite{jarderberg_15} as a learnable module that can be inserted into CNNs and explicitly allows for the spatial manipulation of data within the CNN. In the astronomical context, STNs have only been used by \cite{wu_2019} previously in morphological analysis of radio data. 

In \gampen{}, the STN is upstream of the CNN, where it applies a two-dimensional affine transformation to the input image, and the transformed image is then passed to the CNN. 
%The transformation applied by the STN depends on the specific input image, and the STN 
Each input image is transformed differently by the STN, which 
learns the appropriate cropping during the training of the downstream CNN without additional supervision. As shown in the upper part of Figure \ref{fig:gampen_schematic}, the STN consists of (1) a localization network, (2) a parameterized grid generator, and (3) a sampler, as described below.

\begin{figure*}[htb]
    \centering
    \includegraphics[width
    =\textwidth]{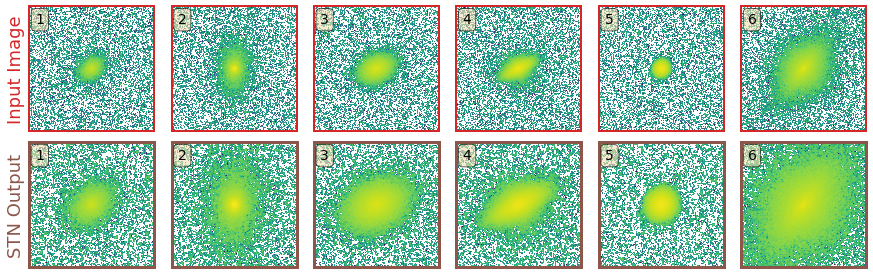}
    \caption{Examples of the transformation applied by the STN to six randomly selected input galaxy images. The top row shows the input galaxy images, and the bottom row shows the corresponding output from the STN. The numbers in the top-left yellow boxes help correspond the output images to the input images. As can be seen, the STN learns to apply an optimal amount of cropping for each input galaxy.}
    \label{fig:stn_examples}
\end{figure*}

The localization network takes the input image, $U$ ($U \in \mathbb{R}^{H \times W \times C}$, with height $H$, width $W$, and $C$ channels), and outputs $\boldsymbol{\theta}$, the six-parameter matrix of the affine transformation, $\mathcal{T}_\theta$, to be applied to the input image.
%\boldsymbol{\theta}4=f_{\mathrm{loc}}(U)$. 
The localization network in the STN is a CNN with two convolutional layers followed by two fully connected layers at the end. 

%Next, the parameterized sampling grid helps to apply the transformation $\mathcal{T}_\theta$ to the input image. To transform the input image, the output pixels are computed by applying a sampling kernel on the input image. 

To perform the transformation, $\mathcal{T}_\theta$, the values of the output pixels are computed by applying a sampling kernel on the input image. As the first step in this process, the parameterized grid generator is used to generate a grid, $G$, of target coordinates, $ G_i =  \left( x_i^t, y_i^t \right) $, forming the output of the STN. For our case, $\mathcal{T}_\theta$ is a 2D affine transformation $A_{\theta}$, and the pointwise transformation is given by

\begin{equation}
\begin{split}
\left(\begin{array}{c}
x_{i}^{s} \\
y_{i}^{s}
\end{array}\right)=\mathcal{T}_{\theta}\left(G_{i}\right) & = \mathrm{A}_{\theta}\left(\begin{array}{c}
x_{i}^{t} \\
y_{i}^{t} \\
1
\end{array}\right) \\
& = \left[\begin{array}{lll}
\theta_{11} & \theta_{12} & \theta_{13} \\
\theta_{21} & \theta_{22} & \theta_{23}
\end{array}\right]\left(\begin{array}{c}
x_{i}^{t} \\
y_{i}^{t} \\
1
\end{array}\right) ,
\end{split}
\label{eq:stn_transformation}
\end{equation}

\noindent
where $ \left(x_i^s,y_i^s\right)$ are the source coordinates in the input image that define the sample points \citep{jarderberg_15}. The transformation shown in Equation\,\ref{eq:stn_transformation} allows for cropping, translation, rotation, and skewing to be applied to the input image. However, the simulated galaxy images in our dataset are already centered, and our primary aim of using the STN is to achieve optimal cropping; thus, we constrain the type of affine transformations allowed by modifying $\mathrm{A}_{\theta}$ such that 

\begin{equation}
\mathrm{A}_{\theta} = \left[\begin{array}{lll}
s & 0 & 0 \\
0 & s & 0
\end{array}\right] .
\label{eq:cropping_transformation}
\end{equation}

\noindent
The localization network predicts the optimal value of $s$ for each input image. As can be seen from Eq. \ref{eq:cropping_transformation}, $s=1$ results in an identity transformation (i.e., the image output by the STN and the input image are the same). For values of $s < 1$, lower fractions of the input image are retained in the output image. For example, when $s=0.7$, $70\%$ of each side (length/width) of the input image is retained in the output image.

Note that although \gampen{}'s STN does not perform rotations, we are able to induce rotational invariance using our training procedure.  Since our simulated training set is very large, it happens to be that there are many galaxies with different position angles, but similar (other) structural parameters.

In the final step, the sampler takes the set of sampling points $\mathcal{T}_\theta(G)$ along with the input image, $U$, to produce the output image, $V$. Each $ \left(x_i^s,y_i^s\right)$ coordinate in $\mathcal{T}_\theta(G)$ defines the spatial location in the input where a bilinear sampling kernel is applied to get the value at a particular pixel in the output image. This can be written as 

\begin{equation}
\begin{split}
V_{i}^{c}=\sum_{n}^{H} \sum_{m}^{W} U_{n m}^{c} & \max \left(0,1-\left|x_{i}^{s}-m\right|\right) \times \\ & \max \left(0,1-\left|y_{i}^{s}-n\right|\right) , 
\end{split}
\label{eq:smapling_v}
\end{equation}

\noindent
where $U_{n m}^{c}$ is the value at location ($n$,$m$) in channel $c$ of the input, and $V_{i}^{c}$ is the output value for pixel $i$ at location $\left(x_i^t,y_i^t\right)$ in channel $c$. To allow the backpropagation of the loss through this sampling mechanism, we can define the gradients with respect to $U$ and $G$. This allows loss gradients to flow back to the sampling grid coordinates and therefore back to the transformation parameter $s$ and the localization network.

Placing the STN upstream in the \gampen{} framework allows the network to learn how to actively transform the input image to minimize the overall loss function during training. Because the transformation we use is differentiable with respect to the parameters, gradients can be backpropagated through the sampling points $\mathcal{T}_\theta(G)$ to the localization network output $\boldsymbol{\theta}$. This crucial property allows the STN to be trained using standard backpropagation along with the downstream CNN, without any additional supervision. 

Figure \ref{fig:stn_examples} shows examples of the transformations applied by the STN of a trained \gampen{} framework to simulated HSC data. As can be seen, the STN learns to apply an optimal amount of cropping for each input galaxy. 

\begin{figure*}[htb]
    \centering
    \includegraphics[width
    =\textwidth]{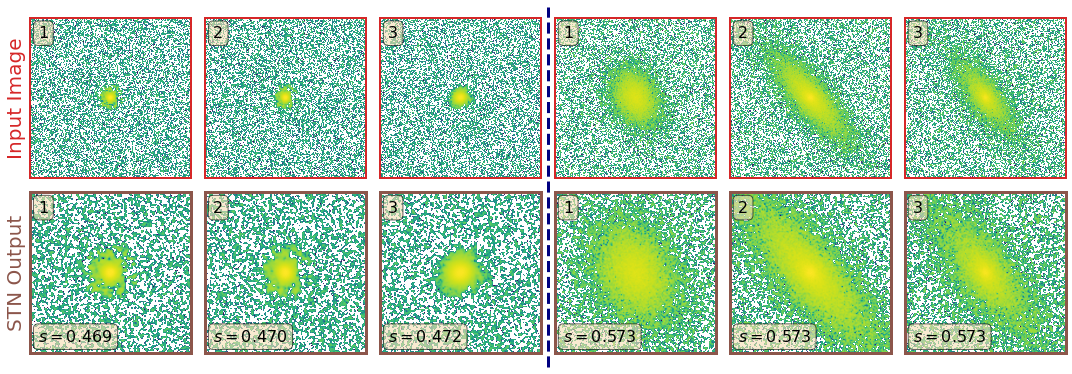}
    \caption{(\textit{Left}): Galaxies in the testing dataset with the lowest values of $s$ (i.e., the most aggressive crops) (\textit{Right}): Galaxies in the testing dataset with the highest values of $s$ (i.e., the least aggressive crops). As can be seen, the STN correctly learns to apply the most aggressive crops to small galaxies; and the least aggressive crops to large galaxies.}
    \label{fig:most_least_crops}
\end{figure*}

To further validate the performance of our STN, we process all images in our testing dataset through the STN module of a trained \gampen{} framework. After that, we sort all the processed images using the value of the parameter $s$ (from Eq. \ref{eq:cropping_transformation}) predicted by the localization network. Higher values of $s$ denote that a more significant fraction of the input image was retained in the output image produced by the STN  (i.e., minimal cropping). In Figure \ref{fig:most_least_crops}, we show the images from our testing dataset with the highest and lowest values of $s$. Figure \ref{fig:most_least_crops} demonstrates that the STN correctly learns to apply the most aggressive crops to smallest galaxies in our dataset, and the least aggressive crops to the largest galaxies.

\begin{figure}[htb]
    \centering
    \includegraphics[width
    =0.45\textwidth]{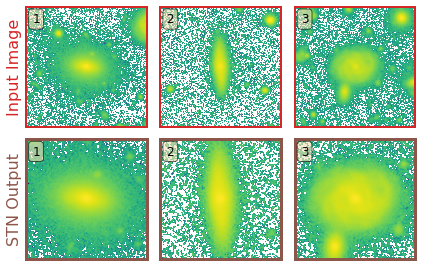}
    \caption{Examples of the transformation applied by a trained STN to real HSC-Wide \textit{g}-band galaxies. The STN helps the downstream CNN to focus on the galaxy of interest at the center of the cutout by cropping out most secondary galaxies present in the input frame.}
    \label{fig:real_galaxies_stn}
\end{figure}

Lastly, in order to demonstrate the purpose of including an STN in \gampen{}, we show its performance on real HSC galaxies. We apply the STN module of a trained \gampen{} framework to three randomly chosen \textit{g}-band galaxies in the HSC-Wide survey with $z \leq 0.25$ and $m_g \leq 23$. Each input image is $40 \times 40$ arcsec. (Note that, for this demonstration, we did not retrain \gampen{} on real galaxies in any way.) The results are shown in Figure\,\ref{fig:real_galaxies_stn}. The STN learns to systematically crop out secondary galaxies in the cutouts and focus on the galaxy of interest at the center of the cutout. At the same time, the STN also correctly applies minimal cropping to the largest galaxies, making sure the entirety of these galaxies remains in the frame. 

\subsection{The Convolutional Neural Network Module} \label{subsec:cnn}

The input image, once transformed by the STN, is passed to the downstream CNN module, as depicted in Figure \ref{fig:gampen_schematic}. This downstream module predicts the posterior distribution of the bulge-to-total light ratio, effective radius, and total flux for each input galaxy. 

The architecture of this downstream CNN is based on the design of VGG-16 \citep{vgg}, a CNN that performed well in the 2014 ImageNet Large Scale Visual Recognition Challenge, wherein different teams compete to classify about 14 million hand-annotated images. The primary feature of the VGG class of networks is that they use tiny convolutional filters combined with significantly deep networks, which have been shown to be highly successful in computer vision. Broadly speaking, \gampen{}'s downstream CNN consists of thirteen convolutional layers, followed by three fully connected layers. The convolutional layers are arranged in five blocks, and in between each block is a max-pooling layer. Note that one of the primary differences between our network and VGG-16 is that all the convolutional layers in \gampen{} are preceded by a dropout layer in order to facilitate the prediction of epistemic uncertainties, as described further in \S \ref{subsec:mcd}. 

The convolutional layers (Conv2D in Fig. \ref{fig:gampen_schematic}) work in unison to identify hierarchies of translational invariant spatial patterns in the images. Each convolutional layer does this by using a collection of $3\times3$ pixel windows (called ``filters"), wherein each filter is a specific pattern that the CNN is looking for in the image. These windows slide around the input to generate a ``response-map" or ``feature-map", which quantifies the presence of the filter's pattern at different locations of the input. Each convolutional layer is preceded by a dropout layer, which is one of the most effective and commonly used regularization techniques that prevent over-fitting by randomly ``dropping" (i.e., setting to zero) several output features of the layer during training. The ``dropout rate" defines the fraction of features that are zeroed out. For \gampen{}, our choice of the dropout rate is guided by calibration of the predicted uncertainties and is described in \S \ref{sec:training}.

The goal of the max-pooling layers (MP in Fig. \ref{fig:gampen_schematic}) is to aggressively down-sample the outputs of the convolutional layer that they follow. Simply speaking, max-pooling is dividing the output of the convolutional layer into a collection of windows and then using the maximum value in each window as the output. Max-pooling can be thought of as a technique for detecting a given feature in a broad region of the image and then throwing away the exact positional information. The intuition is that once a feature has been found, its exact location is not as crucial as its rough location relative to other features. An additional advantage is that by aggressively down-sampling, max-pooling forces successive convolutional layers to look at increasingly large windows as a fraction of the input to the layer. This helps to induce spatial-filter hierarchies. 

Throughout the network, we use the rectified linear unit (ReLU in Fig. \ref{fig:gampen_schematic}) as the activation function, except for the output layer, which is linear. The output of a ReLU unit with input $\mathbf{x}$, weight $\mathbf{w}$, and bias $b$ is given by $\max (0, \boldsymbol{w} \cdot \boldsymbol{x}+b)$. The application of the ReLU activation function makes the network non-linear. 

At the end of the network are three fully connected layers. They use the output of the convolutional layers, denoting the presence of various features in the image, to predict the correct output variables given an image. The output layer predicts nine parameters. Three of these construct the vector of means of the output variables $(\boldsymbol{\hat{\mu}})$, and the remaining six are used to construct the covariance matrix $\boldsymbol{\hat{\Sigma}}$. In \S\,\ref{sec:uncertainties}, we describe more about how these two variables are used to generate the predicted distributions.

Table \ref{tab:network_layers} in the Appendix gives extended descriptions of each \gampen\ layer. 
For more technical details about the various layers and functions described there, we refer the reader to \cite{nielsen,goodfellow_16,chollet_21}.

\section{Prediction of Posteriors} \label{sec:uncertainties}
Traditional CNNs consist of neurons with fixed, deterministic values of weights and biases, resulting in deterministic outputs. However, if the weights in such a network are probability distributions, then the calculation can be defined within a Bayesian framework \citep{denker_91}. Such CNNs can then be used to capture the posterior probabilities of the outputs, resulting in well-defined estimates of uncertainties. The key distinguishing property of the Bayesian approach is marginalization over multiple networks rather than a single optimization run. 

Two primary sources of error contribute to the uncertainties in the parameters predicted by \gampen{}. The first arises from errors inherent to the input imaging data (e.g., noise and PSF blurring), and this is commonly referred to as aleatoric uncertainty. The second error comes from the limitations of the model being used for prediction (e.g., the number of free parameters in \gampen{}, the amount of training data, etc.); this is referred to as epistemic uncertainty. It is important to note that while epistemic uncertainties can be reduced with proper changes to the model (e.g., more training data, more flexible model), aleatoric uncertainties are determined by the input images and thus cannot be reduced. There has been much recent work on how to estimate uncertainties efficiently in deep learning \citep[e.g.,][]{gal_2016,Kendall2017WhatVision,Pawlowski2017ImplicitNetworks,Wilson2020BayesianGeneralization} and some of these techniques have also been applied to astrophysical problems \citep[e.g.,][]{PerreaultLevasseur2017UncertaintiesLensing, Walmsley2020GalaxyLearning, Wagner-Carena2021HierarchicalLensing, Cranmer2021ASystems}. The following two sections describe how we arrange for \gampen\ to estimate both parameter values and their uncertainties.
%as well as model para into the \gampen{} predictions. 

\subsection{Bayesian Implementation of \gampen{} and Epistemic Uncertainties} \label{subsec:mcd}
To create a Bayesian framework while predicting morphological parameters, we have to treat the model itself as a random variable---or more precisely, the weights of our network %, $\boldsymbol{\omega}$, 
must be probabilistic distributions instead of single
%deterministic 
values. For a network with weights, $\boldsymbol{\omega}$, and a training dataset, $\mathcal{D}$, of size $N$ with input images $\left\{\boldsymbol{X}_{1}, \ldots, \boldsymbol{X}_{N}\right\}$ and output parameters $\left\{\boldsymbol{Y}_{1}, \ldots, \boldsymbol{Y}_{N}\right\}$, the posterior of the network weights, $p(\boldsymbol{\omega} \mid \left\{\boldsymbol{X}_{1}, \ldots, \boldsymbol{X}_{N}\right\}, \left\{\boldsymbol{Y}_{1}, \ldots, \boldsymbol{Y}_{N}\right\}) \equiv p(\boldsymbol{\omega} \mid \mathcal{D}) $ represents the plausible network parameters. To predict the probability distribution of the output variable $\boldsymbol{\hat{Y}}$ given a new test image $\boldsymbol{\hat{X}}$, we need to marginalize over all possible weights $\boldsymbol{\omega}$:

\begin{equation}
p(\boldsymbol{\hat{Y}} \mid \boldsymbol{\hat{X}}, \mathcal{D})=\int p(\boldsymbol{\hat{Y}} \mid \boldsymbol{\hat{X}}, \boldsymbol{\omega}) p(\boldsymbol{\omega} \mid \mathcal{D}) d \boldsymbol{\omega} .
\label{eq:out_y_pred}
\end{equation}

In order to calculate the above integral, we need to know $p(\boldsymbol{\omega}\mid\mathcal{D})$, i.e., how likely is a particular set of weights given the available training data, $\mathcal{D}$. Since we have trained only the one model, it does not tell us how likely different sets of weights are. Different approximations have been introduced in order to calculate this distribution, with variational inference \citep{Jordan1999IntroductionModels} being the most popular.

Now, the dropout technique was introduced by \cite{Srivastava2014Dropout:Overfitting} in order to prevent neural networks from overfitting; they temporarily removed random neurons from the network according to a Bernoulli distribution, i.e., individual nodes were set to zero with a probability, $p$, known as the dropout rate. This dropout process can also be interpreted as taking the trained model and permuting it into a different one \citep{Srivastava2014Dropout:Overfitting}.

Using variational inference and dropout, we can approximate the integral in Equation \ref{eq:out_y_pred} as 

\begin{equation}
\int p(\boldsymbol{\hat{Y}} \mid \boldsymbol{\hat{X}}, \boldsymbol{\omega}) p(\boldsymbol{\omega} \mid \mathcal{D}) d \boldsymbol{\omega} \approx \frac{1}{T} \sum_{t=1}^{T} p(\boldsymbol{\hat{Y}} \mid \boldsymbol{\hat{X}}, \boldsymbol{\omega}_t) ,
\label{eq:mcd_final}
\end{equation}

\noindent
wherein we perform $T$ forward passes with dropout enabled and $\boldsymbol{\omega}_t$ is the set of weights during the $t$\textsuperscript{th} forward pass. This procedure is what is referred to as Monte Carlo Dropout. For a detailed derivation of Equation \ref{eq:mcd_final}, please refer to Appendix \ref{sec:ap:mcd_deri}

In order to obtain epistemic uncertainties for \gampen{}, we insert a dropout layer before every weight layer in the network. Each forward pass through \gampen{} samples the approximate parameter posterior. Thus, in order to obtain epistemic uncertainties, we feed every test image $\boldsymbol{\hat{X}}_i$ to the trained \gampen{} framework $T$ times and collect the outputs. In implementing the Monte Carlo Dropout technique, an often-ignored key step is tuning the dropout rate (i.e., the rate at which neurons are set to zero). We discuss the tuning of the dropout rate for \gampen{} in \S \ref{sec:training}.

\vspace{0.8cm}

\subsection{Likelihood Calculation and Aleatoric Uncertainties}
\label{subsec:aleatoric}
Our simulated training data consists of noisy input images by design, but we know the corresponding morphological parameters with perfect accuracy. However, due to the different amounts of noise in each image, the predictions of \gampen{} at test time should have different levels of uncertainties. Thus, in this situation, we want to use a heteroscedastic model -- a model that can capture different levels of uncertainties in its output predictions. We achieve this by training \gampen{} to predict the aleatoric uncertainties. 

As outlined in \S \ref{sec:training}, \gampen{} predicts a multivariate log-normal distribution of output variables for any given input image. Thus, for a given set of network weights $\boldsymbol{\omega}$, the likelihood $p(\mathcal{D} \mid \boldsymbol{\omega})$ is simply the product of the probabilities that the \gampen{} output for each image is drawn from the associated multivariate Gaussian distribution $\mathcal{N}(\boldsymbol{\mu}, \boldsymbol{\Sigma})$ in $\mathbf{R^3}$ , with mean $\boldsymbol{\mu}$ and covariance matrix $\boldsymbol{\Sigma}$.

Although we would like to use \gampen{} to predict aleatoric uncertainties, the covariance matrix, $\boldsymbol{\Sigma}$, is not known {\it a priori}. Instead, we train \gampen{} to learn these values by minimizing the negative log-likelihood of the output parameters for the training set, which can be written as

\begin{equation}
\begin{split}
- \log \mathcal{L}_{VI} \propto  \sum_{n} & \frac{1}{2}\left[\boldsymbol{Y}_{n}-\boldsymbol{\hat{\mu}}_{n}\right]^{\top} \boldsymbol{\hat{\Sigma_n}}^{-1}\left[\boldsymbol{Y}_{n}-\boldsymbol{\hat{\mu}}_{n}\right] \\ 
& + \frac{1}{2} \log [\operatorname{det}(\boldsymbol{\hat{\Sigma_n}})] + \lambda \sum_{i}\left\|\boldsymbol{\omega_{i}}\right\|^{2} .
\end{split}
\label{eq:final_loss_fn}
\end{equation}

where $\boldsymbol{\hat{\mu}}_n$ and $\boldsymbol{\hat{\Sigma}}_n$ are the mean and covariance matrix of the multivariate Gaussian distribution predicted by \gampen{} for an image, $\boldsymbol{X}_n$. $\lambda$ is the strength of the regularization term, and $\boldsymbol{\omega}_i$ are sampled from $q(\boldsymbol{\omega})$. For a detailed derivation of Equation \ref{eq:final_loss_fn}, we refer an interested reader to Appendix \ref{sec:ap:final_loss_deri}.

The covariance matrix here represents the uncertainties in the predicted parameters arising from inherent corruptions to the input or the output data. Note that using the full covariance matrix in Equation \ref{eq:final_loss_fn} instead of just the diagonal terms (i.e., assuming the output variables to be independent), helps \gampen{} to incorporate the structured relationship between the different output parameters. We further outline the effects of this in \S \ref{sec:training}.

\subsection{Practical Implementation Details} \label{subsec:uncertainty_implementation}

In order to predict $\boldsymbol{\mu}$ and $\boldsymbol{\Sigma}$, the final layer of \gampen{} contains nine output nodes (see Fig.\,\ref{fig:gampen_schematic}). Three of these nodes are used to characterize $\boldsymbol{\mu}$. % = \mu_{k=1}^3$, where $k$ represents each output variable. 
Now, although $\boldsymbol{\Sigma}$ is a $3\times3$ matrix, we are able to characterize it with just six parameters due to its special properties. Because $\boldsymbol{\Sigma}$ is a symmetric, positive-definite matrix, we can use the LDL decomposition, a variant of the Cholesky decomposition \citep{cholesky}, to represent 

\begin{equation}
\boldsymbol{\Sigma} = \left(\begin{array}{ccc}
\sigma_{1}^{2} & \sigma_{12} & \sigma_{13} \\
\sigma_{21} & \sigma_{2}^{2} & \sigma_{23} \\
\sigma_{31} & \sigma_{32} & \sigma_{3}^{2}
\end{array}\right)
\label{eq:full_sigma_matrix}
\end{equation}

\noindent
as $\boldsymbol{\Sigma} = \boldsymbol{L}\boldsymbol{D}\boldsymbol{L}^{\top}$, where 

\begin{equation}
\boldsymbol{L} = \left(\begin{array}{lll}
1 & 0 & 0 \\
\sigma_{21} & 1 & 0 \\
\sigma_{31} & \sigma_{32} & 1
\end{array}\right)
\label{eq:form_of_L}
\end{equation}

\noindent
and 

\begin{equation}
D=\left(\begin{array}{ccc}
\sigma_{1}^{2} & 0 & 0 \\
0 & \sigma_{2}^{2} & 0 \\
0 & 0 & \sigma_{3}^{2}
\end{array}\right) .
\label{eq:form_of_D}
\end{equation}

\noindent
Thus, three of \gampen{}'s output nodes are used to predict the off-diagonal elements in Equation\,\ref{eq:form_of_L},
%$\sigma_{i>j\geq1}^{3}\sigma_{i,j}$ 
and three more are used to predict $_{i=1}^3s_i$ where $s_i=\log(\sigma_i^2)$. We predict $s_i$ instead of $\sigma_i^2$ in order to achieve better numerical stability during training. 

The loss function, outlined in Equation \ref{eq:final_loss_fn}, contains the determinant and the inverse of $\boldsymbol{\Sigma}$. Calculation of the determinant and the inverse of a matrix are potentially numerically unstable and slow operations. Thus, in order to achieve the maximum speed possible on our GPUs and for numerical stability, we replace these operations using the Cholesky decomposition outlined above and standard linear algebra. 
That is, $\boldsymbol{\Sigma}^{-1}$ can be written as $\boldsymbol{\Sigma}^{-1}=\left(\boldsymbol{L}^{-1}\right)^{\top} \boldsymbol{D}^{-1} \boldsymbol{L}^{-1}$, where 

\begin{equation}
\boldsymbol{D}^{-1}=\left(\begin{array}{ccc}
\frac{1}{\sigma_{1}^{2}} & 0 & 0 \\
0 & \frac{1}{\sigma_{2}^{2}} & 0 \\
0 & 0 & \frac{1}{\sigma_{3}^{2}}
\end{array}\right)
\label{eq:D_inv}
\end{equation}

\noindent
because $\boldsymbol{D}$ is a diagonal matrix. Because $\boldsymbol{L}$ is a lower triangular matrix, we can also write its inverse as  

\begin{equation}
\boldsymbol{L}^{-1}=(\boldsymbol{I}+\boldsymbol{N})^{-1}= \boldsymbol{I} +\sum_{k=1}^{2}(-1)^{k} \boldsymbol{N}^{k} ,
\label{eq:L_inv}
\end{equation}

\noindent
where $\boldsymbol{I}$ is a $3\times3$ identity matrix and $\boldsymbol{N}$ is a strictly lower triangular and nilpotent matrix such that $\boldsymbol{N} = \boldsymbol{L} - \boldsymbol{I}$.

Finally, we can write the $\log (\operatorname{det}(\boldsymbol{\Sigma}))$ as

\begin{equation}
\begin{split}
\log (\operatorname{det}(\boldsymbol{\Sigma})) 
& = \log (\operatorname{det} (\boldsymbol{L}\boldsymbol{D}\boldsymbol{L}^{\top})) \\
& = \log (\prod_i D_{ii}) \\
& = \sum \log D_{ii} .
\end{split}
\label{eq:log_det_sigma}
\end{equation}

\noindent
By combining Equations \ref{eq:D_inv}, \ref{eq:L_inv}, and \ref{eq:log_det_sigma}, we can calculate the log-likelihood outlined in Equation \ref{eq:final_loss_fn} without having to calculate the inverse or determinant of any matrix, allowing us to fully utilize the capabilities of a GPU and avoiding any numerical instabilities. 

\begin{figure*}[htb]
    \centering
    \includegraphics[width
    =\textwidth]{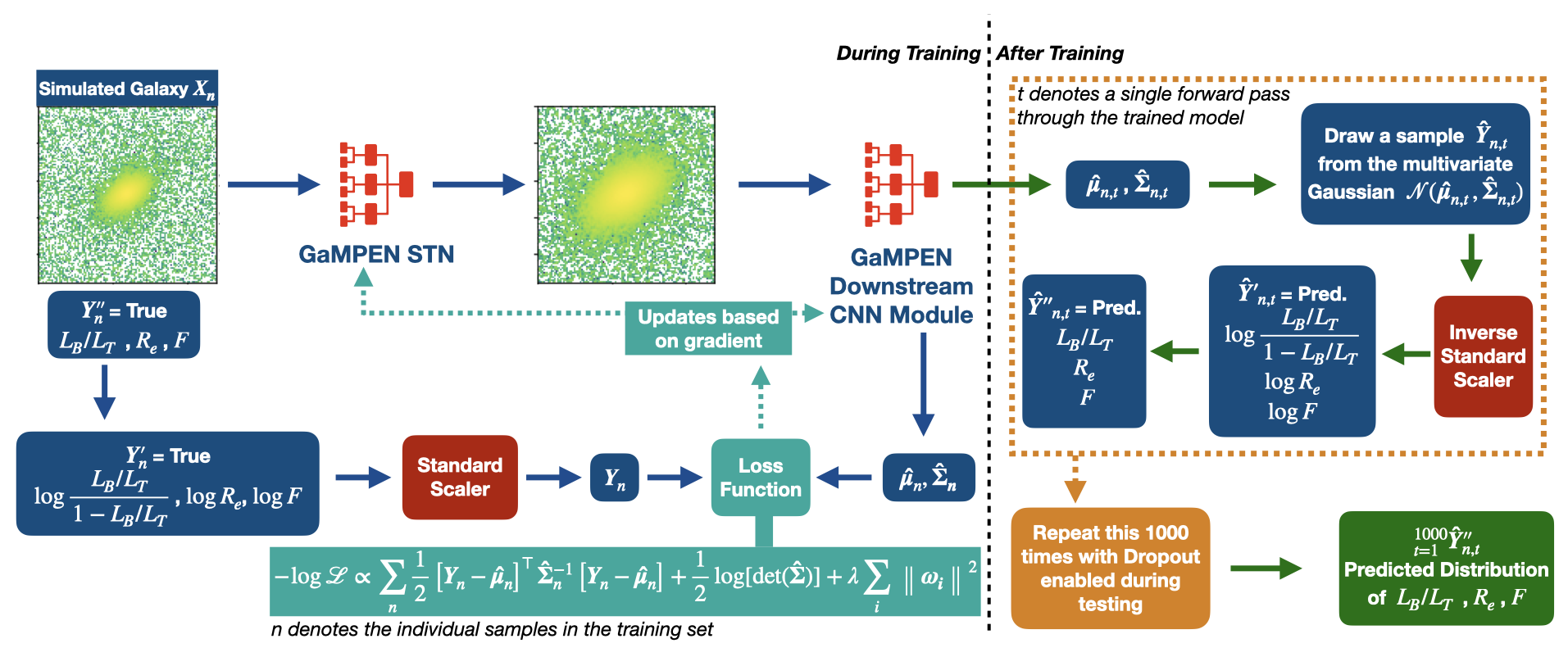}
    \caption{Diagram outlining the training (\textit{left}) and inference (\textit{right}) phases of the \gampen\ workflow. Training consists of feeding 105,000 simulated images (with known parameter values) through the STN and CNN modules, minimizing the loss function (Eqn.\,\ref{eq:final_loss_fn}) using Stochastic Gradient Descent. 
    During this process, we re-scale the variables as described in the text, and return them to the original variable space during inference.
    After the STN+CNN are trained, the inference step consists of 1000 forward passes with dropout enabled for each galaxy image. We draw a sample from the predicted multivariate Gaussian distribution during each forward pass, and the collection of these samples gives us the predicted posterior distribution.}
    \label{fig:workflow}
\end{figure*}

\subsection{Combining Aleatoric and Epistemic Uncertainties}
\label{subsec:combining_uncertainties}
To obtain the posterior distribution of the output variables, we need to combine the aleatoric and epistemic uncertainties. After training a model by maximizing the log-likelihood outlined in Equation \ref{eq:final_loss_fn}, we perform Monte Carlo Dropout. To do this, as outlined in Figure \ref{fig:workflow}, we feed each input image, $\boldsymbol{\hat{X}}_n$, in the test set 1000 times into \gampen{} with dropout enabled. During each iteration, we collect the predicted set of $\left(\hat{\boldsymbol{\mu}}_{n,t},\boldsymbol{\hat{\Sigma}}_{n,t}\right)$ for the $t^{th}$ forward pass. Then, for each forward pass, we draw a sample $\boldsymbol{\hat{Y}_{n,t}}$ from the multivariate normal distribution $\mathcal{N}\left(\boldsymbol{\hat{\mu}}_{n,t},\boldsymbol{\hat{\Sigma}}_{n,t}\right)$. 

The distribution generated by the collection of all 1000 forward passes, $\boldsymbol{\hat{Y}_{n}}$,
%$_{t=1}^{1000}\boldsymbol{\hat{Y}_{n,t}}$ 
represents the predicted posterior distribution for the test image $\boldsymbol{\hat{X}}_n$. The different forward passes capture the epistemic uncertainties, and each prediction in this sample also has its associated aleatoric uncertainty represented by $\boldsymbol{\hat{\Sigma}_{n,t}}$. Thus the above procedure allows us to incorporate both aleatoric and epistemic uncertainties in the prediction of posteriors of morphological parameters by \gampen{}.

\section{Training \gampen } \label{sec:training}
We split the dataset of 150,000 simulated galaxies into training, validation, and testing sets with $70\%$, $15\%$, and $15\%$ of the total sample, respectively. We train \gampen{} using the training set, and set the values of various hyper-parameters (e.g., learning rate, batch size; see below) using the validation set. Finally, we evaluate the trained model on the testing set (which has never been seen before by the network) and report the results in \S \ref{sec:results}.

We pass all the images in the simulated dataset through the arsinh function to reduce the dynamic range of pixel values in the images. This function behaves linearly around zero and logarithmically for large values. Reducing the dynamic range of pixel values has been found to be helpful in neural network convergence \citep[e.g.,][]{zanisi_21,walmsley_decals,tanaka_22}, hence this approach. 

The three output variables that we predict with \gampen{} have quite different ranges, by orders of magnitude. Thus, we re-scale these ground-truth training values before feeding them into the network in order to prevent variables with larger numerical values from making a disproportionate contribution to the loss function. Additionally, we also need to make sure that none of the values predicted by \gampen{} happen to be unphysical; that is, we require all output values to adhere to the following ranges: $0 \leq L_B/L_T \leq 1$; $R_e > 0$; $F > 0$. 

\begin{figure*}[htb]
    \centering
    \includegraphics[width
    =\textwidth]{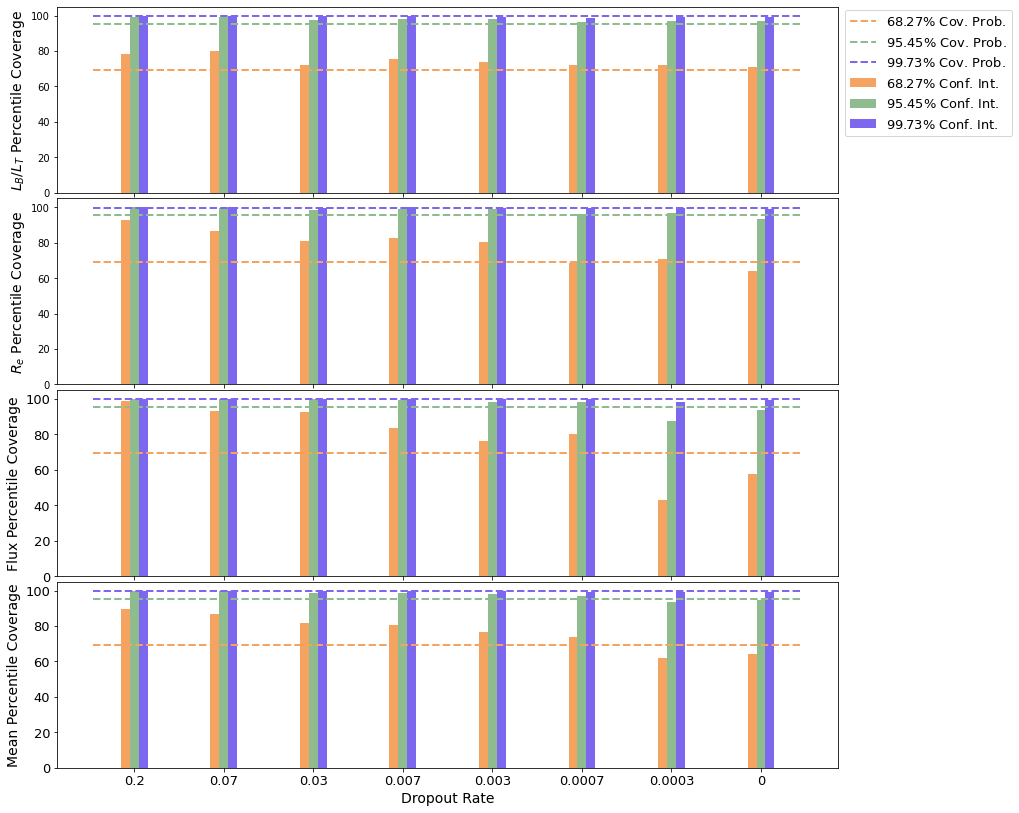}
    \caption{The calculated percentile coverage probabilities for different dropout rates. The top three rows show coverage probabilities for each output variable individually, while the bottom row shows the probabilities averaged over the three variables. The coverage probabilities are defined as the percentage of the total test examples where the  true  value  of  the  parameter  lies  within  a  particular confidence interval of the predicted distribution. A dropout rate of $7\times10^{-4}$ leads to coverage probabilities very close to their corresponding confidence levels.}
    \label{fig:dropout_calibration}
\end{figure*}

Therefore, we first apply the logit transformation to $L_B/L_T$ and log transformations to $R_e$ and $F$: 

\begin{equation}
\boldsymbol{Y_n'} = f''(\boldsymbol{Y}_n'') = \left( \log \frac{L_B/L_T}{1 - L_B/L_T}, \log R_e, \log F \right) ,
\label{eq:transformation_f''}
\end{equation} 

\noindent
where $\boldsymbol{Y}_n'' = [{L_B/L_T,R_e,F}]$ is the set of ground-truth parameters corresponding to the simulated image, $\boldsymbol{X}_n$, and $f''$ is how we will refer to the transformation in Equation\,\ref{eq:transformation_f''}. Note that the uniformity of these transformations allows us to write the likelihood in terms of a multivariate Gaussian distribution. Next we apply the standard scaler to each parameter (calibrated on the training data), which amounts to subtracting the mean value of each parameter and scaling its variance to unity:
%This transformation, which we denote by $f'$, can be written as

\begin{equation}
\boldsymbol{Y}_n = f'(\boldsymbol{Y}_n')\,\,\,\,\,\mathrm{where}\,\,\,\,\,{Y}_{n,i} = \frac{Y'_{n,i} - \overline{Y'_{i}}}{\sqrt{\operatorname{var}(Y'_{i})} ,} 
\label{eq:transformation_f'}
\end{equation}

\noindent
where the $i$ subscript refers to each of the three parameters. Combining the above transformations, $f'$ and $f''$, ensures that all three variables have similar numerical ranges.  

Note that effectively \gampen{} is trained in the $\boldsymbol{Y}_n$ variable space, and the predictions made by \gampen{} are also in this space. Thus, post training, during inference, we need to apply the inverse of the standard scaler function, $f'^{-1}$ (with no re-tuning of the mean or variance), followed by the inverse of the logit and log transformations, $f''^{-1}$, as indicated in Figure\,\ref{fig:workflow}. These final transformations also ensure that the predicted values are all within the physical ranges mentioned earlier.

We train \gampen{} by minimizing the loss function in Equation\,\ref{eq:final_loss_fn} using Stochastic Gradient Descent, wherein we estimate the gradient of the loss function using a mini-batch of training samples and update the network weights and biases accordingly. Calculation of the gradient is done using the back-propagation algorithm, and we refer an interested reader to \citet{rumelhart_88} for details.

The training process involves hyper-parameters that must be chosen: the learning rate (the step-size during gradient descent), momentum (acceleration factor used for faster convergence), strength of L2 regularization ($\lambda$ in the loss function in Eqn.\,\ref{eq:final_loss_fn}), and batch size (the number of images processed before weights and biases are updated).
To choose these hyper-parameters, we trained \gampen{} with a given set of hyper-parameters for forty epochs, then verified convergence by checking whether the value of the loss function and the mean-absolute-error on the validation set had stabilized over at least the last ten epochs. An epoch of training refers to running all of the images in the training set through the network once. We chose final hyper-parameters that resulted in the lowest value for the loss function. This resulted in the following values: Learning Rate, $5\times10^{-7}$; Momentum, 0.99; Strength of L2 regularization $\lambda=10^{-4}$, and Batch Size, 16. The grid of values we used for the hyper-parameter search is as follows:- Learning Rate - $10^{-5}$, $5 \times 10^{-5}$, $10^{-6}$, $5 \times 10^{-6}$, \ldots,  $5 \times 10^{-8}$; Momentum - 0.8, 0.9, 0.95, 0.99; $\lambda$ - $10^{-5}$,$10^{-4}$, \ldots, $10^{-2}$; Batch Size: 8, 16, 32, 64.

One of the most critical adjustable parameters is the dropout rate, as it directly affects the calculation of the epistemic uncertainties (as described in \S \ref{subsec:mcd}). On average, higher dropout rates lead networks to estimate higher epistemic uncertainties. To determine the optimal value for the dropout rate, we trained variants of \gampen{} with dropout rates from 0 to 0.2, all with the same optimized values of momentum, learning rate, and batch size given above. After that, we performed inference using each model as outlined in Figure \ref{fig:workflow}. 

To compare these models, we calculated the percentile coverage probabilities associated with each model, defined as the percentage of the total test examples where the true value of the parameter lies within a particular confidence interval of the predicted distribution. We calculate the coverage probabilities associated with the $68.27\%$, $95.45\%$, and $99.73\%$ central percentile confidence levels, corresponding to the $1\sigma$, $2\sigma$, and $3\sigma$ confidence levels for a normal distribution. For each distribution predicted by \gampen{}, we define the $68.27\%$ confidence interval as the region on the x-axis of the distribution that contains $68.27\%$ of the most probable values of the integrated probability distribution. In order to estimate the probability distribution function from the \gampen{} predictions (which are discrete), we use kernel density estimation, which is a non-parametric technique to estimate the probability density function of a random variable. 

We calculate the $95.45\%$ and $99.73\%$ confidence intervals of the predicted distributions in the same fashion. Finally, we calculate the percentage of test examples for which the true parameter values lie within each of these confidence intervals. An accurate and unbiased estimator should produce coverage probabilities equal to the confidence interval for which it was calculated (e.g., the coverage probability corresponding to the $68.27\%$ confidence interval should be $68.27\%$).

Figure \ref{fig:dropout_calibration} shows the coverage probabilities for the three output parameters individually (top three panels), as well as the coverage probabilities averaged over the three output variables (bottom panel). As can be seen, higher values of the dropout rate lead to \gampen{} over-predicting the epistemic uncertainties, resulting in too high coverage probabilities. In contrast, extremely low values lead to \gampen{} under-predicting the epistemic uncertainties. For a dropout rate of $7\times10^{-4}$, the calculated coverage probabilities are very close to their corresponding confidence levels, resulting in accurately calibrated posteriors. The dropout rate is clearly a variational parameter of \gampen{}, and all the results shown hereafter correspond to a \gampen{} model trained with a dropout rate of $7\times10^{-4}$.

It is important to note that the inclusion of the full covariance matrix in the loss function allowed us to incorporate the relationships between the different output variables in \gampen{} predictions. This allowed us to achieve simultaneous calibration of the coverage probabilities for all three output variables. 
In contrast, using only the diagonal elements of the covariance matrix resulted in substantial disagreement, for a fixed dropout rate, among the coverage probabilities of the different parameters.
%Experiments, assuming the three output variables to be independent, did not result in simultaneous calibrated coverage probabilities. 
Additionally, when we used three different neural networks to predict each output variable, we achieved a poorer overall accuracy. Thus, using the full covariance matrix, facilitated by the linear algebraic tricks outlined in \S \ref{subsec:uncertainty_implementation}, allows \gampen{} to predict accurate, calibrated posteriors.

\section{Results} \label{sec:results}
After training \gampen{} and tuning its hyper-parameters on the training and validation sets, as outlined in \S \ref{sec:training}, we perform inference using the testing set of 22,500 galaxies. 

\subsection{Inspecting the Predicted Posteriors} \label{subsec:predicted_dists}

As outlined in Figure \ref{fig:workflow} and \S \ref{subsec:combining_uncertainties}, during the inference phase, we pass each image in the testing set 1000 times through \gampen{}. Note that due to our use of Monte-Carlo Dropout, each of these forward passes happens through a slightly different network because of how the technique drops out (sets to zero) randomly selected neurons according to a Bernoulli distribution. This technique allows us to effectively factor in the uncertainty about our predictive model into \gampen{} predictions. For each forward pass $t$ and for each test image $\boldsymbol{\hat{X}}_n$, \gampen{} predicts a vector of means $\boldsymbol{\hat{\mu}}_{n,t}$ and a covariance matrix $\boldsymbol{\hat{\Sigma}}_{n,t}$. These two parameters are used to define a multivariate Gaussian distribution $\mathcal{N}(\boldsymbol{\hat{\mu}}_{n,t},\boldsymbol{\hat{\Sigma}}_{n,t})$ from which we draw a sample $\boldsymbol{\hat{Y}}_{n,t}$. These are then processed through two sets of transformations ($f'^{-1}$ and $f''^{-1}$) outlined in \S \ref{sec:training} resulting in the transformed prediction $\boldsymbol{\hat{Y}''}_{n,t}$. The collection of these samples for the 1000 forward passes $_{t=1}^{1000}\boldsymbol{\hat{Y}''}_{n,t}$ represents the posterior distribution predicted by \gampen{} for the test image $\boldsymbol{\hat{X}}_{n}$.

\begin{figure}[htb]
    \centering
    \includegraphics[width
    =0.47\textwidth]{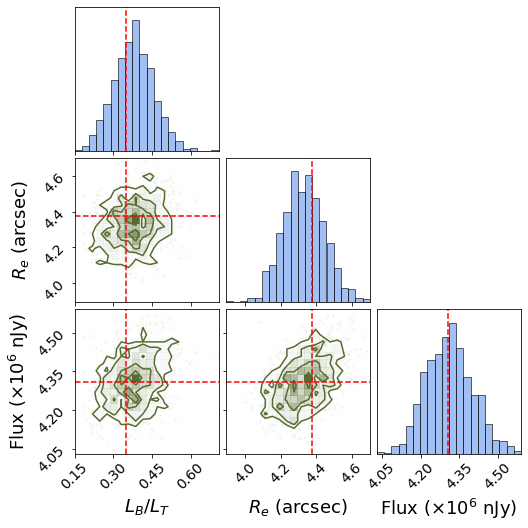}
    \caption{Joint and marginalized probability distributions predicted by \gampen{} for a randomly chosen galaxy in our testing set. The red dotted lines show the true values of the parameters.}
    \label{fig:corner}
\end{figure}

\begin{figure*}[htb]
    \centering
    \includegraphics[width
    =\textwidth]{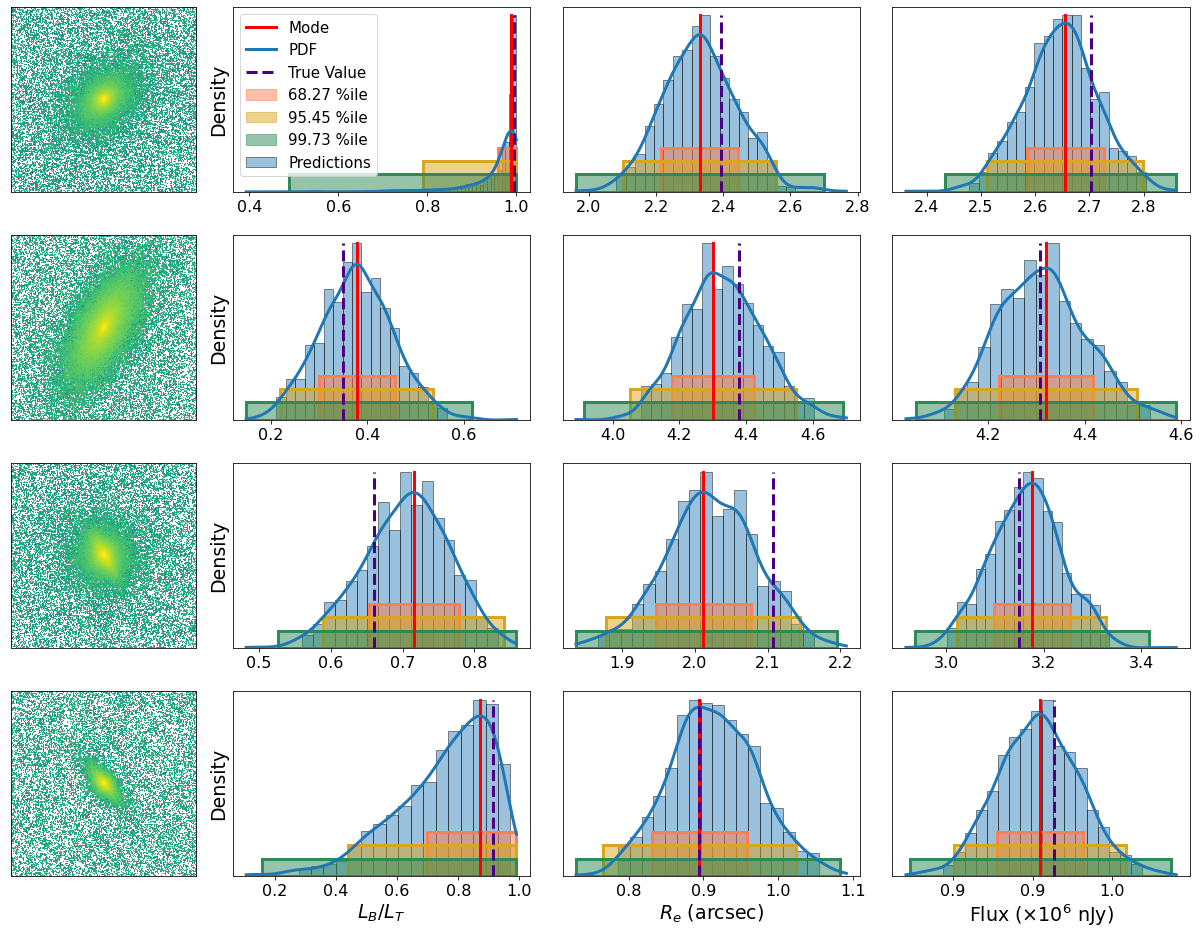}
    \caption{Examples of predicted posterior distributions for four randomly chosen simulated galaxies. The blue shaded histogram shows the predictions from \gampen\, and the blue solid lines show the associated probability distribution functions estimated by kernel density estimation. These are used to calculate the confidence intervals shown in the figure with pink, yellow, and green shading. The mode (red line) shows the most probable value of each morphological parameter. As expected, in most cases, the true value (purple line) lies within the $68.27\%$ confidence interval.}
    \label{fig:example_pred_dists}
\end{figure*}

Using the above process, we extract the joint probability distribution of all the output parameters for each of the 22,500 galaxies in our test set. Figure \ref{fig:corner} shows the two-dimensional joint distributions of the output parameters, as well the marginalized distributions, for a randomly chosen galaxy in our test set. 
%The red dotted lines in the figure show the true values of the parameters. 
The same galaxy %for which the joint distributions are plotted in Figure \ref{fig:corner} 
is shown in the second row of Figure \ref{fig:example_pred_dists},
which illustrates the predicted posterior distributions for a few more cases.
%Note that for this specific galaxy, the true values of the different parameters do lie in the densest parts of the probability density space predicted by \gampen{}. In order to show the predicted distributions for a few more cases and also to visually demonstrate the calculation of the probability density function as well as the confidence intervals, we show the marginalized histograms for a few more galaxies in Figure \ref{fig:example_pred_dists}.
%In addition to the marginalized distributions, 
Figure \ref{fig:example_pred_dists} also shows the image of each galaxy %in the leftmost column. 
at the left.
As expected, all the predicted distributions are unimodal, smooth, and resemble Gaussian or truncated Gaussian distributions. For each predicted distribution, the figure also shows the parameter space regions that contain $68.27\%$, $95.45\%$, and $99.73\%$ of the most probable values of the integrated probability distribution. We use kernel density estimation to estimate the probability distribution function (PDF; shown by a blue line in the figure) from the predicted values. The mode of this PDF is what we refer to as the predicted value when calculating residuals. In the figure, for most cases, the true value lies within the most probable $68.27\%$ percentile region. We also visually inspected the distributions predicted by \gampen{} for $\sim200$ galaxies to ensure that there were no systematic or catastrophic errors (e.g., substantial errors for a specific parameter only, or bi-modal or irregular distributions for specific kinds of galaxies, etc.). 

By design, \gampen{} predicts only physically possible values. This is especially apparent in the $L_B/L_T$ column of rows 1 and 4 of Figure \ref{fig:example_pred_dists}. Note that to achieve this, we do not artificially truncate these distributions. Instead, we use the inverse of the logit transformation on the prediction space of \gampen{} as outlined in \S \ref{sec:training}. This ensures that the predicted $L_B/L_T$ values are always between 0 and 1. Similarly, we also ensure  that the $R_e$ and $F$ values predicted by \gampen{} are positive through appropriate transformations. 

\begin{figure*}[htb]
    \centering
    \includegraphics[width
    =\textwidth]{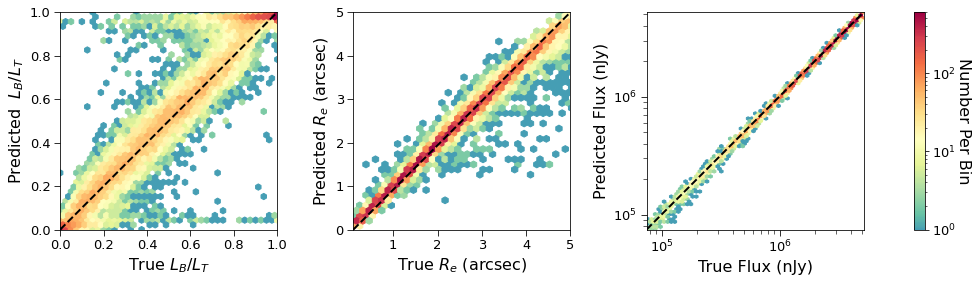}
    \caption{The true values of the galaxy parameters plotted against the most probable values predicted by \gampen{}. The black dashed line marks the $y=x$ diagonal on which perfectly recovered parameters should lie. The color of each hexagon corresponds to the number of galaxies it contains, as indicated by the colorbar at right.}
    \label{fig:pred_true}
\end{figure*}

\subsection{Evaluating the Accuracy of \gampen{}} \label{subsec:residuals}

In \S \ref{subsec:predicted_dists}, we explored the predicted distributions for a handful of cases, where the true values of the parameters mostly lay within the densest parts of the probability distribution predicted by \gampen{}. In order to evaluate the accuracy of \gampen{}, we now report summary statistics that outline the framework's performance on the entire testing set.

\begin{deluxetable}{c|ccc}[htbp]
%\tablenum{2}
\tablecaption{Coverage Probabilities on the Test Set  \label{tab:coverage_probs}}
\tablecolumns{4}
\tablehead{
\colhead{Parameter } & \colhead{$68.27\%$} & \colhead{$95.45\%$} & \colhead{$99.73\%$} \\
\colhead{Name} & \colhead{Conf. Level} & \colhead{Conf. Level} & \colhead{Conf. Level} 
}
\startdata
    \hline
    \hline
    $L_B/L_T$ & $71.8\%$ &  $96.9\%$ &  $98.9\%$\\
    $R_e$ &  $68.1\%$ &  $95.9\%$ &  $98.3\%$ \\
    $F$ &  $78.7\%$ &  $98.2\%$ &  $99.9\%$\\
    \hline
    Mean & $72.9\%$ &  $97.0\%$ & $99.0\%$\\
\enddata
\tablecomments{The coverage probabilities are defined as the percentage of the total test samples where the  true  value  of  the  parameter  lies  within  a particular confidence interval of the predicted distribution.}
\end{deluxetable}

In Table \ref{tab:coverage_probs}, we report the coverage probabilities that \gampen{} achieves on the test set. 
% (i.e., the percentage of test examples where the true values of the parameter lie within a particular confidence threshold of the predicted distribution, see \S\,\ref{sec:training}). 
In the ideal situation, they would perfectly mirror the confidence levels; that is, $68.27\%$ of the time, the true value would lie within $68.27\%$ of the most probable volume of the predicted distribution.
(Note that in \S\,\ref{sec:training} we tuned the dropout rate so they coincide over the validation set, whereas Table\,\ref{tab:coverage_probs} is calculated on the testing set.) 
%Note that we tuned the dropout rate based on the coverage probabilities on the validation set in \S \ref{sec:training}, whereas Table\,\ref{tab:coverage_probs} is calculated on the testing set.
%Table \ref{tab:coverage_probs} demonstrates that 
Clearly, \gampen{} produces well calibrated and accurate posteriors, 
%: although the coverage probabilities are not perfectly aligned for the all three parameters, they are 
consistently close to the claimed confidence levels.
%to allow the posteriors to be used for most practical purposes. 
Additionally, we note that even for the flux, for which the coverage probabilities are most discrepant, the uncertainties predicted by \gampen{} are in any case overestimates (i.e., conservative). %Thus, \gampen{} is not over-confident for any of the confidence thresholds. 
If \gampen{} were used in a scenario that requires perfect alignment of coverage probabilities, users could employ techniques such as importance sampling \citep{importance_sampling} on the distributions predicted by \gampen{}. 
%In future work, we plan to try techniques such as deep ensembles \citep{deep_ensembles} to see whether these lead to even better calibrated coverage probabilities. 

\begin{figure*}[htb]
    \centering
    \includegraphics[width
    =\textwidth]{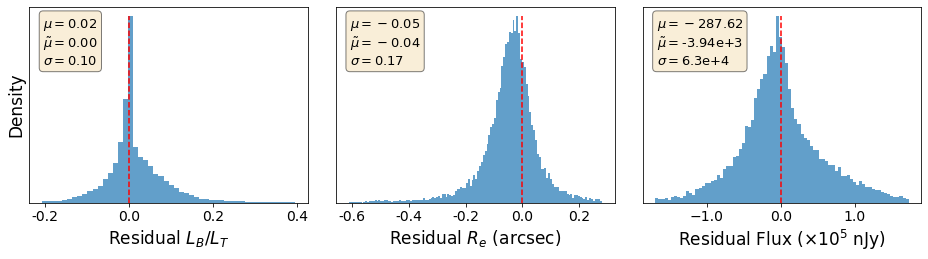}
    \caption{Histograms of residuals for all galaxies in the testing set. We define the residuals as the difference between the true value and the most probable value predicted by \gampen{}. The dashed vertical line represents $x = 0$, denoting cases with perfectly recovered parameter values. 
    The mean ($\mu$), median ($\tilde{\mu}$), and standard deviation ($\sigma$) of each residual distribution are listed in each panel.
    %The box in the upper left corner mentions the mean ($\mu$), median ($\tilde{\mu}$), and the standard deviation ($\sigma$) of each residual distribution.
    }
    \label{fig:residual_hists}
\end{figure*}

%We would like to note here again that predicting the joint posterior distribution allowed us to incorporate the structured relationship between the different output parameters within \gampen{}, allowing us to achieve simultaneous calibration of the coverage probabilities for all the three output variables. In experiments where we assumed the three output variables to be independent of each other didn't result in simultaneous calibration of coverage probabilities for all the three variables. 

Having defined the overall percentage of cases where the true values are within particular confidence levels of the predicted distributions, we now quantify the difference between the most probable values of the predicted parameters (i.e., modes of these predicted distributions) and the true values. 

Figure\,\ref{fig:pred_true} shows the most probable values predicted by \gampen{} for the testing set versus the true values, %. The distribution of galaxies in this predicted v/s true space is shown 
%using a hexbin color plot, where the space is divided into 
in hexagonal bins of roughly equal size, with the number of galaxies represented according to the colorbar on the right. % side of Figure \ref{fig:pred_true}. 
We have used a logarithmic colorbar to visualize even small clusters of galaxies in this plane. 
Most galaxies are clustered around the line of equality, showing that the most probable values of the distributions predicted by \gampen{} closely track the true values of the parameters. 

The middle panel of Figure\,\ref{fig:pred_true} shows a small bias (note that the color scale is logarithmic) towards low predictions of $R_e$, especially for true $R_e > 4$ arcsec. Features like this have been seen before in other machine learning studies and are typically referred to as an “edge effect” -- that is, sometimes the model performs poorly at the edges of the parameter space on which it was trained. Here, since $R_e=5$ arcsec is the largest radius present in our training data, for some galaxies with $R_e$ close to 5 arcsec, the network is hesitant to predict the highest value it has ever seen and predicts a slightly lower value. This results in the small bias toward lower predicted values of $R_e$. However, note that despite this effect for a small number of galaxies; even at a  large radius, \gampen{} accurately estimates $R_e$ for the large majority of galaxies. Among some other larger deviations evident in Figure\,\ref{fig:pred_true}, are predictions near the limits of $L_B/L_T$. We explore this further below. 

In Figure \ref{fig:residual_hists}, we show the residual distribution for the three parameters predicted by \gampen{}. We define the residual for each parameter as the difference between the most probable predicted value and the true value, i.e., $\operatorname{Mode}(\boldsymbol{\hat{Y}_n}) - \boldsymbol{Y_n}$. The box in the upper left corner gives 
the mean ($\mu$), median ($\tilde{\mu}$), and standard deviation ($\sigma$) of each residual distribution. 
All three distributions are normally distributed (verified using the Shapiro Wilk test), and have $\mu \sim \tilde{\mu} \sim 0$.
The \gampen\ prediction of the bulge-to-total ratio is, in $\sim68.27\%$ of cases, within 0.1 of the true value.
The typical error in effective radius is $0.17$ arcsec.
Typical uncertainties in the flux are at the 0.1-1\% level.
%typical fluxes in our dataset are two-three orders of magnitude higher than these values. 
% The values of the reported standard deviations help us to understand the typical size of the errors in the values predicted by \gampen{}. For example, while predicting $L_B/L_T$, in $\sim68.27\%$ of cases, the error in the predicted value of $L_B/L_T$ is $\leq0.1$. The corresponding typical errors for $R_e$ and $F$ are $0.17$ arcsec, and $6.3\times10^4$ nJy respectively. 

\begin{figure*}[htb]
    \centering
    \includegraphics[width
    =\textwidth]{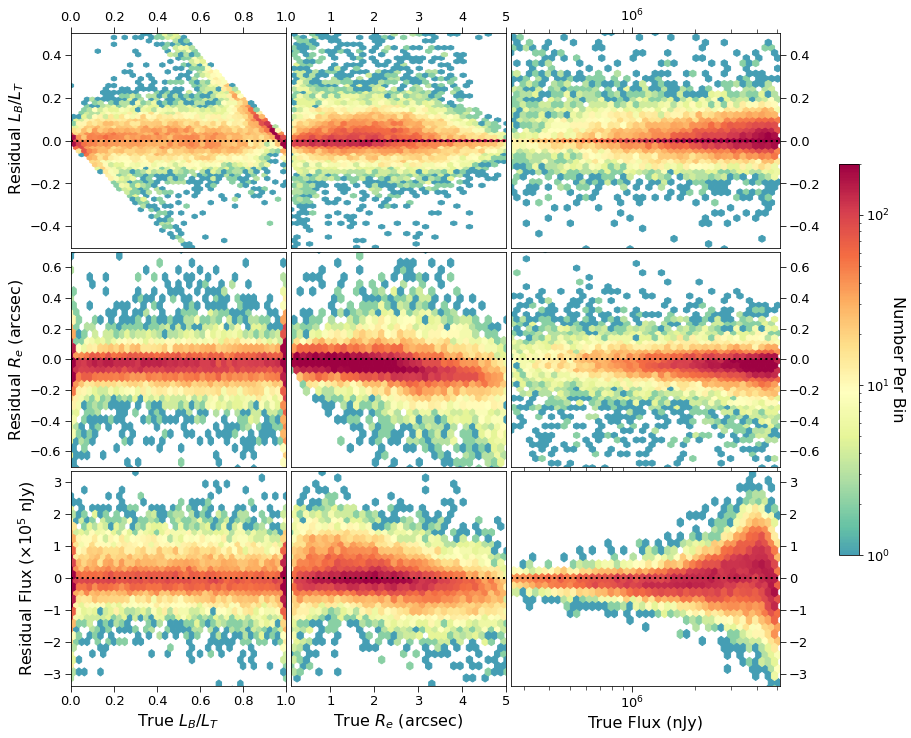}
    \caption{Residuals of \gampen{} predicted parameter values plotted against the true values. The residual for each parameter is defined as the difference between the most probable predicted value and the true value, i.e., $\operatorname{Mode}(\boldsymbol{\hat{Y}_n}) - \boldsymbol{Y_n}$. The color of each hexagonal bin corresponds to the number of galaxies it contains, as shown by the colorbar on the right. The black dotted line ($y=0$) represents %the ideal scenario of 
    perfectly recovered parameters.}
    \label{fig:2d_residual_hists}
\end{figure*}

\begin{figure*}[htb]
    \centering
    \includegraphics[width
    =\textwidth]{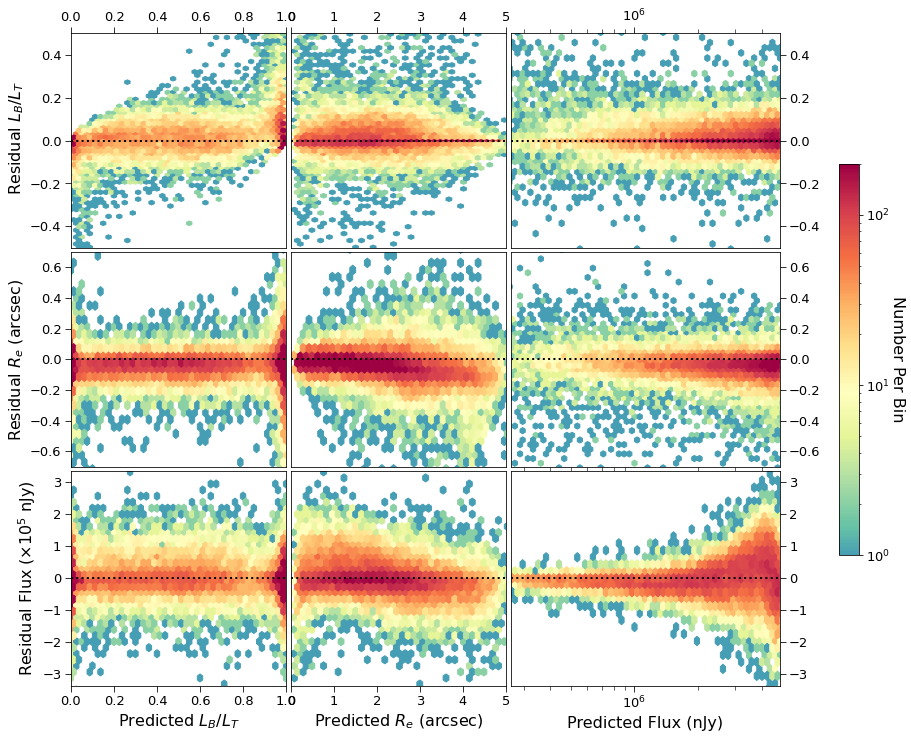}
    \caption{Residuals of the output parameters plotted against the predicted values. This figure allows us to assign quality labels to \gampen{} predictions (e.g., flagging parameters that are unreliable) based on the output values. See \S\,\ref{subsec:cuts} for details.}
    \label{fig:2d_residual_hists_pred}
\end{figure*}

Although Figures \ref{fig:pred_true} and \ref{fig:residual_hists} indicate the overall accuracy of \gampen{}, they do not reveal how those errors depend on location in the parameter space. This is critical information as this enables us to potentially ignore predictions for regions of parameter space that have large errors (according to the validation set).
Figure \ref{fig:2d_residual_hists} shows the residuals for the three output parameters plotted against the true values. As in Figure \ref{fig:pred_true}, we have split the parameter space into hexagonal bins and used a logarithmic color scale to denote the number of galaxies in each bin. The purpose of this plot is to identify regions of parameter space where \gampen{} performs especially well or badly, so that, in the future, we can flag predictions in these regions as ``very secure" or ``unreliable". Note that because we are performing the test here on simulated galaxies, we have access to the ground-truth values. However, in a scenario where \gampen{} is being used on real galaxies which have not been morphologically studied before, we won't have access to the ground-truth values, and any such cuts on the X-axis would need to be made based on the values predicted by \gampen{}. Thus, we created Figure \ref{fig:2d_residual_hists_pred}, where we replaced the X-axis with the predicted values of the parameters instead of the true values. 

\begin{figure}[hbt]
    \centering
    \includegraphics[width
    =0.45\textwidth]{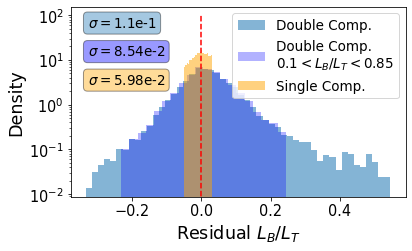}
    \caption{Histograms of $L_B/L_T$ residuals shown separately for single component galaxies, all double component galaxies, and double component galaxies with $0.1 < L_B/L_T < 0.85$. The standard deviation ($\sigma$) for each distribution is also shown in the top left. The dashed vertical line represents $x = 0$, denoting cases with perfectly recovered $L_B/L_T$. The apparent hard cutoffs in the distributions of the single component, and the restricted range double-component galaxies arise from the fact that the y-scale is logarithmic. We have verified that when plotted on a linear scale, the apparent hard cutoffs disappear.}
    \label{fig:residual_hist_bt_sep}
\end{figure}

\begin{figure*}[htb]
    \centering
    \includegraphics[width
    =0.9\textwidth]{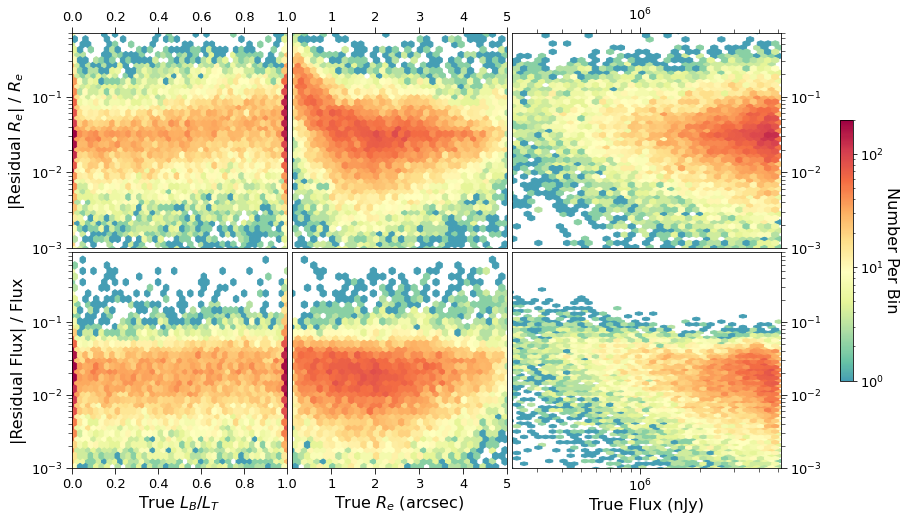}
    \caption{Fractional residuals for the effective radius and flux plotted against their corresponding true values. Note that since we are plotting the absolute values, the ideal situation of perfectly recovered parameters is at the bottom of each panel. The right two panels show that both the residuals increase for fainter galaxies, while the top-middle panel shows that the radius residuals increase for smaller galaxies.}
    \label{fig:2d_residual_hists_norm}
\end{figure*}

In both Figures \ref{fig:2d_residual_hists} and \ref{fig:2d_residual_hists_pred}, for most of the panels, the large majority of galaxies are clustered uniformly around the black dashed line, $y = 0$, which denotes the ideal case of perfectly recovered parameters.% with the residuals being equal to zero. 

There are a few other notable features in these two figures. In the top left panel, the $L_B/L_T$ residuals are highest near the limits of $L_B/L_T$. This is another manifestation of the edge-effect mentioned earlier, wherein sometimes machine learning algorithms perform poorly at the edges of the parameter space on which they were trained.  To delve deeper, we looked at the $L_B/L_T$ residuals separately for double and single component galaxies, as shown in Figure \ref{fig:residual_hist_bt_sep}. 
For single-component galaxies, the typical $L_B/L_T$ residual ($\sigma=0.06$) is roughly half as large as for double-component galaxies ($\sigma=0.11$);
%(i.e., with both bulge and disk components). 
among the latter, the residuals are especially high when $L_B/L_T > 0.85$ or $L_B/L_T < 0.1$. In other words, accurately determining $L_B/L_T$ is challenging when both a bulge and a disk are present, and becomes even more difficult when one component strongly dominates the other. %This is not unexpected since precisely measuring the ratio of the fluxes of two galaxy components is inherently inaccurate if there are two galaxy components and one component completely dominates over the other. 
Larger residuals in the predictions near the limits of $L_B/L_T$ leads to the features seen in the top left panel of both figures.

This edge effect also results in the top and bottom streaks seen in the left panel of Figure \ref{fig:pred_true}. Given the logarithmic  color bar used in this figure, note that most of the galaxies in the upper streak have true $L_B/L_T > 0.75$ and those in the bottom streak have true $L_B/L_T < 0.25$. For these cases, when one component completely dominates over the other component, precisely determining $L_B/L_T$ is challenging. For some of the galaxies with $0.25 > \mathrm{True}\,\,L_B/L_T > 0.75$, \gampen{} assigns almost the entirety of the light to the dominant component, resulting in the streaks in the left panel. We use a parameter transformation to mitigate this edge effect, as described in \S\,\ref{subsec:cuts}. 

In the top-middle panels of Figures \ref{fig:2d_residual_hists} and \ref{fig:2d_residual_hists_pred}, there is a slight broadening of the residuals at low values of the effective radius. This result also makes sense: smaller galaxies are challenging to analyze for any image processing algorithm. Somewhat surprising features appear in the panels showing residuals of effective radius (mid-row, mid-column) and flux (bottom-row, right-column). The $R_e$ residuals increase in magnitude (with a bias towards negative values) toward increasing values of $R_e$, and the residual flux also grows rapidly with higher flux values. However, these increases are simply the result of the increasing numerical values of the parameters. To show this, Figure \ref{fig:2d_residual_hists_norm} plots the dimensionless fractional $R_e$ and $F$ residuals (note that $L_B/L_T$ is inherently dimensionless), using their absolute values, such that the ideal scenario (i.e., zero residuals) is at the bottom of each panel instead of in the middle. 
In this presentation, both features noted above not only disappear but reverse.
For small values of effective radius, $R_e < 1.0$ arcsec, there is an increase in the magnitude of the residuals.
Similarly, the right two panels show that the residuals of $R_e$ and $F$ are systematically higher for faint galaxies, $F< 10^6$ nJy.% A somewhat surprising feature presents itself in the panel showing the residuals of $R_e$ v/s $R_e$ (mid-row; mid-column) and in the panel displaying residuals of the flux v/s the flux (bottom-row; right-column). The $R_e$ residuals seem to increase in magnitude (with a bias towards -ve values) at high values of $R_e$, while the residual flux also grows rapidly with higher flux values. However, it is essential to note that these increases in residuals might be trivially connected to the high numerical values of the parameters. In order to account for this, we plot the dimensionless fractional $R_e$ and $F$ residuals instead of their numerical values in Figure \ref{fig:2d_residual_hists_norm} (note that $L_B/L_T$ is dimensionless inherently). Another difference is that in Figure \ref{fig:2d_residual_hists_norm}, we plot the absolute values, and thus the ideal scenario (i.e., zero residuals) is at the bottom of each panel instead of being in the middle. In Figure \ref{fig:2d_residual_hists_norm}, both the features noticed earlier not only disappear; but reverse. For the residuals of $R_e$ v/s $R_e$ (top row; middle column), we see that for values of $R_e < 1.0$ arcsec, there is an increase in the magnitude of the residuals. Similarly, the right two panels show that the residuals of $R_e$ and $F$ are systematically higher for galaxies fainter than $10^6$ nJy.

In other words, \gampen{} systematically becomes less accurate at predicting the radii of galaxies when their sizes become comparable to the seeing of the HSC-Wide Survey (\textit{g}-band median FWHM $\sim0.85\arcsec$). Similarly, \gampen{} finds it more challenging to predict the sizes and fluxes of fainter galaxies, just as one would expect.
%. This is in line with the fact that fainter and smaller galaxies are inherently difficult to work with using any analysis technique. 
With our previously published classification network, \gamornet{} \citep{Ghosh2020GalaxyGalaxies}, we observed a similar reduction in prediction accuracy for smaller and fainter galaxies. Figures \ref{fig:2d_residual_hists}, \ref{fig:2d_residual_hists_pred}, and \ref{fig:2d_residual_hists_norm} help quantify the errors in \gampen{} predictions in different regions of parameter space. These will be essential in order to interpret results appropriately when applying \gampen{} to real galaxies. 
%We will also explore in \S \ref{subsec:cuts} how transforming some of the predicted values can help us reduce the magnitude of the residuals. 

\vspace{0.8cm}

\subsection{Inspecting the Predicted Uncertainties} \label{subsec:uncertainties}
The primary advantage of a Bayesian ML framework like \gampen{} is its ability to predict the full posterior distributions of the output parameters instead of just point estimates. Thus, we would expect such a network to inherently produce wider distributions (i.e., larger uncertainties) in regions of the parameter space where residuals are higher. Here we delineate regions of the parameter space for which \gampen{} predicts broader distributions and we see that these generally coincide with those that have the largest residuals (Figs.\,\ref{fig:2d_residual_hists}, \ref{fig:2d_residual_hists_pred}, \ref{fig:2d_residual_hists_norm}).  

\begin{figure*}[htb]
	\centering
	\includegraphics[width
	=\textwidth]{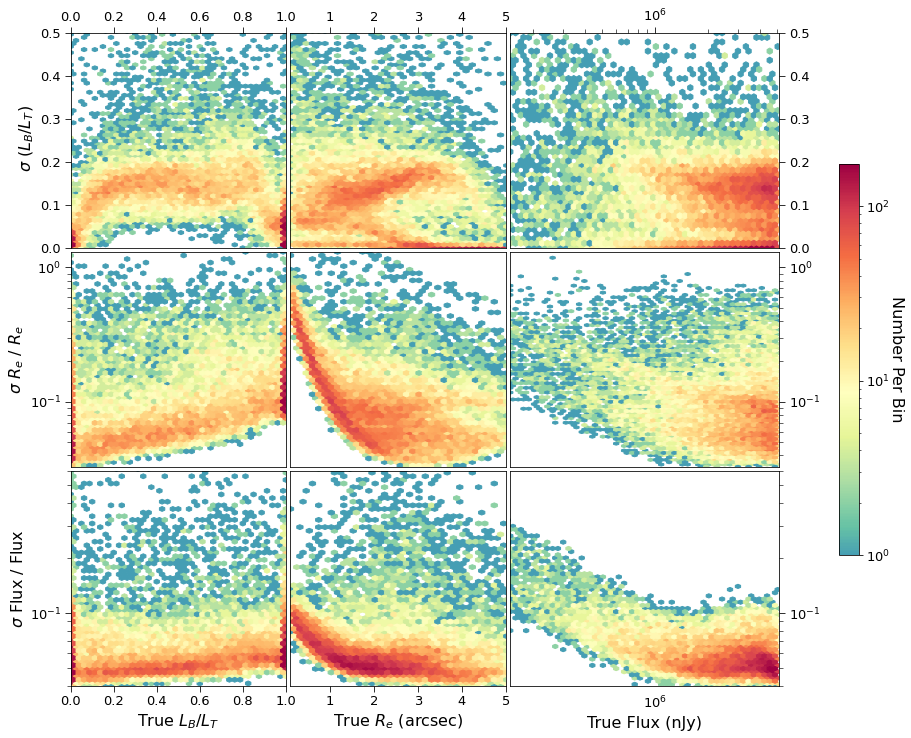}
	\caption{Uncertainties predicted by \gampen{} for each parameter plotted against the true values. The $\sigma$ for each parameter is defined as the width of the $68.27\%$ confidence interval. %i.e., the parameter interval that contains $68.27\%$ of the most probable values;
	Note that we plot fractional uncertainties for radius and flux in order to make the y-axis dimensionless for all three rows.}
	\label{fig:2d_uncertainty_hists_norm}
\end{figure*}

\begin{figure*}[htb]
	\centering
	\includegraphics[width
	=\textwidth]{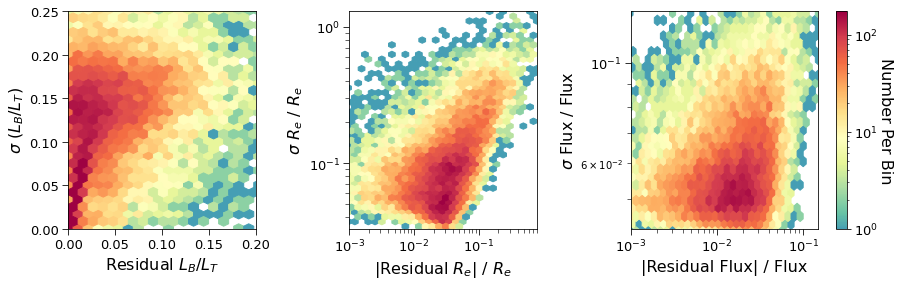}
	\caption{Uncertainties (widths of the $68.27\%$ confidence intervals)  predicted by \gampen{} for each parameter  versus the corresponding residuals (predicted mode minus true value). Fractional uncertainties and residuals are plotted for radius and flux in order to make all the quantities dimensionless. The trend in all three cases is that \gampen{}-estimated uncertainties increase for cases where its predictions are less accurate. The coverage probabilities reported on the test set (Table \ref{tab:coverage_probs}) confirm that the predicted uncertainties are well-calibrated and correspond well to the quoted confidence intervals.}
	\label{fig:uncer_resi}
\end{figure*}

Figure \ref{fig:2d_uncertainty_hists_norm} shows the uncertainties for the three predicted parameters plotted against the true values of the different parameters. 
We define the uncertainty predicted for each parameter as the width of the $68.27\%$ confidence interval (i.e., the parameter interval that contains $68.27\%$ of the most probable values of the predicted distribution; see Fig.\,\ref{fig:example_pred_dists}). 
The lower two panels have been normalized, so that all three panels show dimensionless fractional uncertainties.
%Note that to make the uncertainties for all three parameters dimensionless, we plot fractional uncertainties instead of numerical uncertainties .

The uncertainties in the predicted values of the radius increase sharply for galaxies with $R_e < 2$ arcsec and/or $F < 10^6$ nJy. Similarly, the uncertainties in flux increase for galaxies with $R_e < 1$ arcsec and/or $F < 10^6$ nJy. 
%Both these results show that the uncertainties predicted by \gampen{} perfectly align with our expectation that any model should be typically more uncertain in its predictions at smaller values of $R_e$ and $F$. These regions also correlate with regions where the residuals were higher, as shown in Figure \ref{fig:2d_residual_hists_norm}. 
This aligns perfectly with what we expect: the sizes and fluxes of small galaxies and/or faint galaxies are not well constrained.
Just as the residuals for \gampen{} predictions were larger for small and/or faint galaxies (Figs.\,\ref{fig:2d_residual_hists}, \ref{fig:2d_residual_hists_pred}, \ref{fig:2d_residual_hists_norm}), the uncertainties predicted by \gampen{} are also larger for these galaxies.
%compensates for the higher residuals obtained for smaller and fainter galaxies by predicting higher uncertainties in these regions. Therefore, despite having higher uncertainties in some areas of the parameter space, distributions predicted by \gampen{} are still accurate and well-calibrated, as shown in Table \ref{tab:coverage_probs}.

The top left panel of Figure \ref{fig:2d_uncertainty_hists_norm} shows that \gampen{} is reasonably certain of its predicted %$L_B/L_T$ values
bulge-to-total ratio across the full range of values but appears slightly more certain when $L_B/L_T \leq 0.2$ or $L_B/L_T \geq 0.8$. It turns out that the smaller uncertainties at the limits correspond to the single-component galaxies, while for the double-component galaxies, the edge effect is less pronounced, in agreement with the residuals observed in Figure \ref{fig:residual_hist_bt_sep}.
%when a galaxy is completely disk dominated or bulge dominated, but is more uncertain when %$0.2 \leq L_B/L_T \leq 0.8$. 
%Therefore, \gampen{} assigns higher uncertainties to its $L_B/L_T$ predictions when there is a comparable amount of light in both the disk and bulge components. This is because such galaxies are intrinsically more challenging to classify as disk or bulge-dominated than galaxies where one component is significantly stronger than the other.
Not surprisingly, the predicted uncertainty in $R_e$ decreases with decreasing values of $L_B/L_T$ (i.e., galaxies with more dominant disks, which are on average larger than bulge-dominated galaxies in our simulation sample). 

%The top left panel of Figure \ref{fig:2d_residual_hists_norm} also shows a slight trend in the same direction. We believe that this result is connected to the fact that disk-dominated galaxies have a higher maximum radius limit in our simulations, as shown in Table \ref{tab:sim_para}. This causes the average radius of disk-dominated galaxies to be higher than bulge-dominated galaxies. Since the uncertainty in the predicted value of $R_e$ decreases at higher values of $R_e$, it results in disk-dominated galaxies having lower $\sigma R_e/R_e$ compared to bulge-dominated galaxies. 

In Figure \ref{fig:uncer_resi}, we further assess the estimated uncertainties by investigating their relation to the measured residuals. Note that while the uncertainties represent the widths of the central $68.27\%$ confidence intervals, the residuals are the difference between the modes of the predicted distributions and the true values. Thus, we do not expect the two values to be linearly correlated; rather, on average, \gampen{} should predict larger uncertainties for galaxies with larger residuals, as is seen in all three panels of the figure. 
%Note that the streak of galaxies with extremely small $L_B/L_T$ residuals is dominated by single-components galaxies (as shown in Figure 15) for which \gampen{} correctly predicts correspondingly smaller uncertainties.
According to a Spearman's rank correlation test \citep[see][for more details]{spearman}, there is a positive correlation between the residuals and uncertainties for all three variables, and the null hypothesis of non-correlation can be rejected at extremely high significance ($p < 10^{-200}$). 
%We found the uncertainties and residuals to be positively correlated for all three variables with a correlation coefficient of 0.62, 0.46, and 0.28 for $L_B/L_T$, $R_e$, and flux respectively. This shows that for all three output variables, in general, \gampen{} produces higher uncertainties for cases where the residuals are higher -- resulting in the calibrated coverage probabilities shown in Table 2. This outlines the primary advantage of using a Bayesian framework like \gampen{} -- even in situations where the network is not perfectly accurate, it is able to predict the right level of precision, allowing its predictions to be reliable and well-calibrated.

The results shown in this section outline the primary advantage of using a Bayesian framework like \gampen{} -- even in situations where the network is not perfectly accurate, it is able to predict the right level of precision, allowing its predictions to be reliable and well-calibrated.

\subsection{Qualitative Transformation of \gampen{} Predictions} \label{subsec:cuts}

Given that we know \gampen{} residuals are higher for certain regions of the parameter space, we explore how using only qualitative labels in those regions (instead of quantitative predictions) affects the overall residual values. The labeling is informed by the results of \S \ref{subsec:residuals} and the labels are assigned by us based on the parameter values predicted by \gampen{}.
The labels are applied based on the predicted values of \gampen{} because we will not have access to true values of the parameters when applying \gampen{} to previously unanalyzed real galaxies. This is crucial given that when we apply \gampen{} to real data, techniques like this will provide us practical tools to deal with predictions in regions of the parameter-space where we know \gampen{} to be less accurate. 

\begin{figure*}[htb]
    \centering
    \includegraphics[width
    =\textwidth]{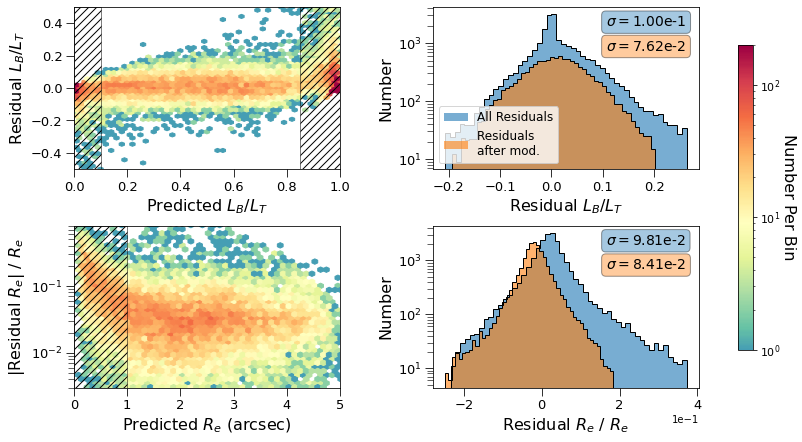}
    \caption{The left panels show the residuals for bulge-to-total light ratio and radius plotted against their predicted values. The black dashed regions show the parameter-space where we replace the quantitative predictions with qualitative flags. Each corresponding histogram on the right shows the distribution of residuals before and after the transformation of output values.}
    \label{fig:exclude_zones}
\end{figure*}

\begin{figure}[htb]
    \centering
    \includegraphics[width
    =0.4\textwidth]{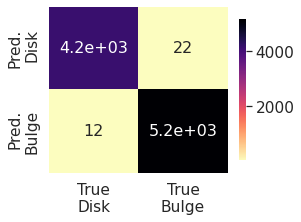}
    \caption{Confusion matrix between the labels we assign when \gampen\ predicts extreme bulge-to-total ratios, $L_B/L_T <0.1$ or $>0.85$, and their true $L_B/L_T$ values. % for those galaxies. 
    The number in each block shows how many galaxies correspond to that panel, resulting in an overall accuracy $>99\%$.
    }
    \label{fig:confusion_matrix}
\end{figure}

For the bulge-to-total ratio, we retain \gampen's numerical predictions for $0.1 < L_B/L_T < 0.85$, but label the more extreme galaxies as ``highly bulge-dominated" ($L_B/L_T \geq 0.85$) or ``highly disk-dominated" ($L_B/L_T \leq 0.1$).
%This is essentially saying that we are confident in \gampen{}'s exact determination of $L_B/L_T$ in the zone $0.1 < L_B/L_T < 0.85$. However, when \gampen{} predicts values  $L_B/L_T \geq 0.85$ or $\leq 0.1$, although we cannot trust the exact predicted value with a high degree of certainty, we can strongly assert that the galaxy is a strongly bulge-/disk-dominated galaxy respectively.%For $L_B/L_T$, we transform the predictions by \gampen{} such that we retain all quantitative predictions in the range $0.1 < L_B/L_T < 0.85$; but transform all values of $L_B/L_T \geq 0.85$ to a flag---``highly bulge-dominated" and all values of $L_B/L_T \leq 0.1$ to a flag---``highly disk-dominated". This is essentially saying that we are confident in \gampen{}'s exact determination of $L_B/L_T$ in the zone $0.1 < L_B/L_T < 0.85$. However, when \gampen{} predicts values  $L_B/L_T \geq 0.85$ or $\leq 0.1$, although we cannot trust the exact predicted value with a high degree of certainty, we can strongly assert that the galaxy is a strongly bulge-/disk-dominated galaxy respectively. 
The top left panel of Figure \ref{fig:exclude_zones} shows the two labeled regions (black-shaded grid), which is where the residuals are highest. 
%As is evident from the figure, we choose the regions for the transformation based on the high residual values. 
The right panel of the top row shows the residual distributions including and excluding the extreme cases. As indicated by the standard deviation (top right corner), removing these extreme cases eliminates the largest errors in the predicted values of $L_B/L_T$. We also checked the accuracy of our assigned labels, and show the confusion matrix in Figure \ref{fig:confusion_matrix}. From this, we calculate the net accuracy of our extreme $L_B/L_T$ labels to be $\gtrapprox99\%$. 

We apply similar labels to small predicted values of the effective radius. As shown in the bottom row of Figure\,\ref{fig:exclude_zones}, we flag galaxies with $R_e < 1.0$ arcsec with the label ``galaxy with $R_e < 1$ arcsec" in place of the exact numerical value. %Similar to what we observed for $L_B/L_T$, the said transformation 
This reduces the typical error for $R_e$, as shown in the histogram on the right. We calculate the accuracy of this label to be $\sim97\%$.

Thus, replacing \gampen's quantitative predictions in certain small regions of the parameter space with qualitative flags results in a reduction of the typical residuals as well as highly accurate qualitative predictions.

%\subsection{Comparing \gampen{} predictions to values obtained by model fitting}
%\begin{itemize}
%    \item Description of model fitting procedure
%    \item Overall comparison of results
%    \item Conclusions you can draw from this
%\end{itemize}

%\subsection{Interpretability of the CNN?} \label{subsec:interpretability}
%\begin{itemize}
%    \item Do you want to do this?
%    \item Adapt some of the results that you achieved previously for this new %regression network
%    \item Or confess that you did this for a previous classification version, but %are now releasing this. 
%\end{itemize}

\section{Discussion \& Conclusions} \label{sec:conclusions}
In this work, we introduced the Galaxy Morphology Posterior Estimation Network (\gampen{}), a machine learning framework that can estimate posterior distributions for a galaxy's bulge-to-total light ratio, effective radius, and flux. Although \gampen{} was trained to estimate these specific parameters, it can be adapted by users easily to predict other/additional morphological parameters (e.g., axis ratio, position angle, etc.). One important consideration while choosing how many parameters to predict using a single \gampen{} framework is that the number of terms in the covariance matrix will increase as $\mathcal{O}(n^2)$, where n is the number of output variables. Although the computation time will increase at a much less steep rate, the exact nature of the increase will depend on the specifics of the hardware being used.

We trained \gampen{} on two  NVIDIA Tesla P100/V100 GPUs with each training run taking about $\sim 12 - 16$ hours. \gampen{} is designed to use multiple GPUs during training and using more GPUs can reduce this training time even further. Our hyperparameter search required $\sim 30$ runs. Given that we expect $\sim 100,000$ images to always be enough to train \gampen{}, our framework can easily be trained on other datasets within a similar reasonable timescale. Once trained, it takes \gampen{} less than a millisecond to process each input galaxy image. Thus, to predict distributions ($\sim 1000$ inference runs for each image) for a million galaxies on two GPUs, \gampen{} needs $\sim 5$ days of runtime. Therefore, \gampen{} can be used to process data from future large surveys like LSST, NGRST, and Euclid within a reasonable timescale.

Training and testing \gampen{} on galaxies simulated to match Hyper Suprime-Cam Wide \textit{g}-band $z<0.25$ data, we found excellent agreement between the coverage probabilities and the corresponding confidence thresholds (Table \ref{tab:coverage_probs}). 
This demonstrates that \gampen{} predicted posterior distributions are calibrated and accurate.

To account for both aleatoric and epistemic uncertainties in \gampen{} predictions, we incorporated the full covariance matrix in our loss function and used the Monte Carlo Dropout technique. Using the covariance matrix also allowed us to incorporate the structured relationships among the output parameters into \gampen{} predictions. This made it possible to achieve the simultaneous calibration of the posteriors of all three output variables (Fig. \ref{fig:dropout_calibration}). In order to incorporate the covariance matrix in the loss function, we used the Cholesky decomposition and a set of linear algebraic tricks (\S \ref{subsec:uncertainty_implementation}).

The typical values of errors in \gampen\ predictions are $0.10$, $0.17$ arcsec {($\sim 7\%$)}, and $6.3\times10^4$ nJy ($\sim 1\%$), for $L_B/L_T$, $R_e$, and $F$ respectively. The error in \gampen{} predictions of $R_e$ increases when $R_e < 1$ arcsec (i.e., when $R_e$ becomes comparable to the seeing of the HSC-Wide survey) and/or $F < 10^6$ nJy. Galaxies fainter than $10^6$ nJy also result in higher flux residuals. These trends result from the inherent challenge in analyzing small and faint galaxies. \gampen{} accounts for these high residuals by correctly predicting higher uncertainties in $R_e$ and $F$ for smaller and fainter galaxies. In other words, \gampen{} predicts broader distributions in regions where it is less precise.

The residuals in \gampen{} predictions of $L_B/L_T$ are high for $L_B/L_T \sim 0$ and $\sim 1$. We demonstrate that by applying qualitative labels for $0.1 \geq L_B/L_T \geq 0.85$ instead of quantitative values, we can reduce the typical error in $L_B/L_T$ to $0.076$. The produced qualitative labels (in the regions with high residuals) have extremely high accuracies of $\gtrapprox99\%$. Similarly, by labeling predictions for $R_e < 1$ arcsec, we achieve a similar reduction in the typical $R_e$ residual, and the produced labels are highly accurate. Thus, the qualitative transformation of the output values gives us tools to deal with regions of the parameter space where residuals in \gampen{} predictions are high.

It is difficult to accurately compare \gampen{}'s performance to existing morphology parameter estimation pipelines (such as GALFIT \citep{galfit}, GIM2D \citep{gim2d}, or \citet{Tuccillo2018DeepFitting}'s neural network) primarily due to the fact that none of them estimate Bayesian posteriors for the predicted parameters. GALFIT and GIM2D do include analytical estimates of errors, but \citet{haussler_07} found that both these algorithms severely underestimate the true uncertainties by an extremely large factor ($\geq 70\%$ for most galaxies). In contrast, \gampen{}'s predicted uncertainties are well-calibrated and accurate ($<5\%$ deviation). Although \gampen{}'s predictions are best used and interpreted in a probabilistic context, we compare below \gampen{}'s residuals (assuming the most probable value to be the predicted value) to the residuals achieved by other frameworks.

\citet{meert_13} used GALFIT to fit two-component light profiles to simulated Sloan Digital Sky Survey \citep[SDSS; ][]{sdss_1_2_des} galaxies and found the typical $R_e$ error to be $10\%$ and the typical magnitude error to be 0.075 mag. \gampen{} achieves a typical $R_e$ error of $\sim7\%$ and a typical magnitude error of 0.051 mag. \citet{haussler_07} used single-component fits to analyze simulated galaxies in the Hubble Space Telescope (HST) Galaxy Evolution from Morphology and SEDs \citep[GEMS; ][]{gems} survey  using GALFIT and GIM2D. They found the typical error in magnitude to be 0.05 mag using GALFIT and 0.10 mag using GIM2D. They found the typical ratio between the predicted and true $R_e$ to be $0.98\pm0.06$ using GALFIT and $1.01 \pm 0.11$ using GIM2D. The same value for \gampen{} is $0.98\pm0.08$. \citet{Tuccillo2018DeepFitting} used a CNN to obtain predictions for the parameters of a single component \sersic{} fit using simulations of HST Cosmic Assembly Near-Infrared Deep Extragalactic Legacy Survey \citep[CANDELS; ][]{candels_1} galaxies  and reported the degree of regression accuracy defined as

\begin{equation}
R^{2}=1-\frac{\sum_{i}^{n}\left(y_{i}-f_{i}\right)^{2}}{\sum_{i}^{n}\left(y_{i}-\bar{y}\right)^{2}}
\end{equation}

where $f_i$ is the predicted value of the true variable $y_i$ and $\bar{y}$ is the mean over all n samples. The $R^2$ for magnitude and $R_e$ were reported to be 0.997 and 0.972 respectively. The authors also analyzed the same galaxies using GALFIT and found the corresponding $R^2$ to be 0.983 and 0.877. The $R^2$ achieved by \gampen{} for magnitude and $R_e$ are 0.998 and 0.980 respectively. While \citet{haussler_07} and \citet{Tuccillo2018DeepFitting} did not have estimates of $L_B/L_T$ (as they used single-component fits), \citet{meert_13} did not report their residuals for $L_B/L_T$. Thus, it was not possible to compare \gampen{}'s $L_B/L_T$ residuals with these previous works. Although none of the above represent absolutely equivalent comparisons, they indicate that \gampen{}'s prediction accuracy is comparable to the most popular state-of-the-art morphology prediction tools.

\gampen{} contains a Spatial Transformer Network that enables it to crop the input image automatically. We demonstrated that \gampen{} does this based on galaxy size, without any need for specific instruction. % therein (without it being engineered to do so). 
Because the transformation is differentiable, loss gradients can be backpropagated, and thus the STN can be trained along with the rest of the framework without any additional supervision. The STN in \gampen{} will empower us to apply it to future large datasets over a broad range of redshifts without having to worry about optimal cutout sizes.  

Although in recent years there has been a significant increase in the use of CNNs for morphological determination, \gampen{} is the first machine learning framework that can robustly estimate posterior distributions of multiple morphological parameters. \gampen{} is also the first application of an STN to optical imaging in astronomy. 

By testing \gampen{} on simulated HSC \textit{g}-band galaxies, where we have access to robust ground-truth values, we demonstrated its effectiveness in recovering morphological parameters and we quantified errors/uncertainties in \gampen{} predictions across different regions of the parameter space. 

Note that, for this work, we trained and tested \gampen{} on single-band images. However, we have tested and verified that both the CNN and STN in \gampen{} can be easily adapted to intake an arbitrary number of channels, with each channel being a different band. To obtain separate morphological parameters for each band, the number of output parameters would need to be increased appropriately. We will perform a detailed evaluation of \gampen{}'s performance on multi-band images in future work.

In this work, we performed a thorough analysis of \gampen{}'s performance using HSC $z < 0.25$ simulations. However, \gampen{} can be applied to a wide variety of other datasets -- including real HSC images, imaging from other ground and space-based observatories as well as higher redshift data. However, in order to apply \gampen{} to real data, one would need to perform appropriate transfer-learning (i.e., fine-tuning the simulation trained \gampen{} models using a small amount of data from the application dataset). We refer an interested reader to \citet{Ghosh2020GalaxyGalaxies}, where we performed transfer-learning and demonstrated the application of our classification framework, \gamornet{}, to SDSS $z\sim0$ and CANDELS $z\sim1$ data.
    
Just like other image analysis methods, we expect \gampen{}'s performance to change based on the quality of the images being used (e.g., the pixel-scale of the survey, the noise, the redshift of the object). We will explore how \gampen{} performs in each of these above situations in future work.

\acknowledgments

The authors would like to thank the anonymous referee for their insightful, encouraging, and extremely thorough comments about our manuscript.  Their constructive criticism has greatly assisted us in improving the manuscript, extending the discussion section, and making it more accessible to readers.

This material is based upon work supported by the National Science Foundation under Grant No. 1715512

CMU and AG would like to acknowledge support from the National Aeronautics and Space Administration via ADAP Grant 80NSSC18K0418. 

AG would like to acknowledge support received from the Yale Graduate School of Arts \& Sciences through the Dean's Emerging Scholars Research Award.

AG would like to acknowledge computing grants received through the Amazon Cloud Credits for Research Program and the Yale Center for Research Computing (YCRC) Resarch Credits Program. AG would also like to acknowledge computing support from YCRC and Yale Information Technology Services staff members and scientists. 

ET acknowledges support from FONDECYT Regular 1190818 and 1200495, ANID grants CATA-Basal AFB-170002, ACE210002, and FB210003, and Millennium Nucleus NCN19\_058.

The Hyper Suprime-Cam (HSC) collaboration includes the astronomical communities of Japan and Taiwan, and Princeton University. The HSC instrumentation and software were developed by the National Astronomical Observatory of Japan (NAOJ), the Kavli Institute for the Physics and Mathematics of the Universe (Kavli IPMU), the University of Tokyo, the High Energy Accelerator Research Organization (KEK), the Academia Sinica Institute for Astronomy and Astrophysics in Taiwan (ASIAA), and Princeton University. Funding was contributed by the FIRST program from Japanese Cabinet Office, the Ministry of Education, Culture, Sports, Science and Technology (MEXT), the Japan Society for the Promotion of Science (JSPS), Japan Science and Technology Agency (JST), the Toray Science Foundation, NAOJ, Kavli IPMU, KEK, ASIAA, and Princeton University. 

This paper makes use of software developed for the Large Synoptic Survey Telescope. We thank the LSST Project for making their code available as free software at  \href{http://dm.lsst.org}{http://dm.lsst.org}.

The Pan-STARRS1 Surveys (PS1) have been made possible through contributions of the Institute for Astronomy, the University of Hawaii, the Pan-STARRS Project Office, the Max-Planck Society and its participating institutes, the Max Planck Institute for Astronomy, Heidelberg and the Max Planck Institute for Extraterrestrial Physics, Garching, The Johns Hopkins University, Durham University, the University of Edinburgh, Queen’s University Belfast, the Harvard-Smithsonian Center for Astrophysics, the Las Cumbres Observatory Global Telescope Network Incorporated, the National Central University of Taiwan, the Space Telescope Science Institute, the National Aeronautics and Space Administration under Grant No. NNX08AR22G issued through the Planetary Science Division of the NASA Science Mission Directorate, the National Science Foundation under Grant No. AST-1238877, the University of Maryland, and Eotvos Lorand University (ELTE) and the Los Alamos National Laboratory.

Based, in part, on data collected at the Subaru Telescope and retrieved from the HSC data archive system, which is operated by Subaru Telescope and Astronomy Data Center at National Astronomical Observatory of Japan.

\clearpage

\appendix

\section{Early Data Access}\label{sec:ap:data_access}
Currently, we are applying \gampen{} to real data, and will make the source code public in the Fall of 2022, along with documentation, tutorials, and trained models. Readers interested in using \gampen{} before the full public release can access the source code and trained models of \gampen{} by emailing the corresponding author of this paper. 

\vspace{10pt}
The public data release for \gampen{} will be hosted at the following two locations:

\begin{itemize}
    \item \href{http://www.ghosharitra.com/}{http://www.ghosharitra.com/}
    \item \href{http://www.astro.yale.edu/aghosh/}{http://www.astro.yale.edu/aghosh/}
\end{itemize}

\section{Extended Derivation for Bayesian Implementation of \gampen{}} \label{sec:ap:mcd_deri}

In variational inference, the posterior, $p(\boldsymbol{\omega} \mid \mathcal{D})$ in Equation \ref{eq:out_y_pred}, is replaced by an approximate variational distribution with an analytic form $q(\boldsymbol{\omega})$. Now, Equation \ref{eq:out_y_pred} can be written as

\begin{equation}
p(\boldsymbol{\hat{Y}} \mid \boldsymbol{\hat{X}}) \approx \int p(\boldsymbol{\hat{Y}} \mid \boldsymbol{\hat{X}}, \boldsymbol{\omega}) q(\boldsymbol{\omega}) d \boldsymbol{\omega} .
\label{eq:out_y_pred_vi}
\end{equation}

The choice of the variational distribution is arbitrary. 
One such choice, introduced by \cite{gal_2016}, involves dropping different neurons from some layers in order to assess the impact on the model. 
The dropout technique was introduced by \cite{Srivastava2014Dropout:Overfitting} in order to prevent neural networks from overfitting; they temporarily removed random neurons from the network according to a Bernoulli distribution, i.e., individual nodes were set to zero with a probability, $p$, known as the dropout rate. 

In the variational application, we use dropouts to interrogate the model. 
Specifically, 
if $p_{i}$ is the probability of a neuron being turned off, and
$\left[z_{i, j}\right]_{j=1}^{J_{i-1}}$ is a vector of length $J_{i-1}$ containing the Bernoulli-distributed random variables for unit $j=1, \ldots, J_{i-1}$ in the $(i-1)^{th}$ layer with probabilities $p_i$, then

\begin{equation}
%\begin{aligned}
\boldsymbol{\omega}_{i} =\boldsymbol{M}_{i} \cdot \operatorname{diag}\left(\left[z_{i, j}\right]_{j=1}^{J_{i-1}}\right) ,
%\end{aligned}
\label{eq:bernoulli}
\end{equation}

\noindent
where $\boldsymbol{M}_i$ is the $J_i \times J_{i-1}$ matrix of variational parameters to be optimized. 

Thus, sampling from $q(\boldsymbol{\omega})$ is now equivalent to using dropouts on a set of layers, with weights $\boldsymbol{M}$ %corresponding to each 
(i.e., $\boldsymbol{M}_i$ for the $i^{th}$ layer). 
We perform inference on the trained network by 
% Once the network has been trained, we can perform inference by 
approximating Equation \ref{eq:out_y_pred_vi} with a Monte Carlo integration: 

\begin{equation}
\int p(\boldsymbol{\hat{Y}} \mid \boldsymbol{\hat{X}}, \boldsymbol{\omega}) q(\boldsymbol{\omega}) d \boldsymbol{\omega} \approx \frac{1}{T} \sum_{t=1}^{T} p(\boldsymbol{\hat{Y}} \mid \boldsymbol{\hat{X}}, \boldsymbol{\omega}_t) ,
\end{equation}

\noindent
wherein we perform $T$ forward passes with dropout enabled and $\boldsymbol{\omega}_t$ is the set of weights during the $t$\textsuperscript{th} forward pass.

\section{Extended Derivation of the Loss Function} \label{sec:ap:final_loss_deri}

As outlined in Equation \ref{eq:out_y_pred}, we seek the most likely set of model parameters given our training data, %$\mathcal{D}$, 
i.e., we maximize 

\begin{equation}
p(\boldsymbol{\omega} \mid \mathcal{D}) \propto p(\mathcal{D} \mid \boldsymbol{\omega})p(\boldsymbol{\omega}) .
\label{eq:bayes_rule}
\end{equation}

In Equation\,\ref{eq:bayes_rule}, $p(\boldsymbol{\omega})$ is the prior on the neural networks weights. The weight prior here is unimportant and what matters is the prior induced on the output parameters of \gampen{}. And as outlined above, we use an uninformative multivariate Gaussian prior to induce an uninformative prior on the output. Please refer to \cite{wilson_20} for a detained discussion on priors in Bayesian deep learning. 

For a regression task using a standard CNN, wherein the network outputs predictions  $\hat{\boldsymbol{Y}}_{n}\left(\boldsymbol{\hat{X}}_{n}, \boldsymbol{\omega}\right)$ for true values $\boldsymbol{Y}_{n}$, one popular choice is to minimize the squared-error loss function $\sum_{n} \left\|\boldsymbol{Y}_n-\boldsymbol{\hat{Y}}_n\left(\boldsymbol{\hat{X}}_n, \boldsymbol{\omega}\right)\right\|^{2}$, where the sum over $n$ denotes a sum over the training set. However, in contrast to the traditional approach, for each new test image $\boldsymbol{\hat{X}}$, \gampen{} needs to predict the parameters of a multivariate Gaussian distribution, $\mathcal{N}(\boldsymbol{\mu}, \boldsymbol{\Sigma})$. 

Now, as discussed in \S \ref{subsec:mcd}, we replace $p(\boldsymbol{\omega} \mid \mathcal{D})$ in Equation \ref{eq:out_y_pred} with an approximating variational distribution $q(\boldsymbol{\omega})$. This is performed by minimizing their Kullback-Leibler (KL) divergence, a measure of the similarity between two distributions. Since minimizing the KL divergence is equivalent to maximizing the log-evidence lower bound,

\begin{equation}
\log \mathcal{L}_{\mathrm{VI}}= \int q(\omega) \log p( \left\{\boldsymbol{Y}_{n=1}^{N}\right\} \mid \left\{\boldsymbol{X}_{n=1}^N\right\}, \boldsymbol{\omega}) d \boldsymbol{\omega}  - \mathrm{KL}(q(\omega) \| p(\omega)) .
\label{eq:log_likelihood_vi}
\end{equation}

\noindent
The first term in Equation \ref{eq:log_likelihood_vi} is the log-likelihood for the output parameters for the training set, and as shown in \cite{gal_2016}, the KL term can be approximated as an $L_2$ regularization. Therefore, Equation \ref{eq:log_likelihood_vi} can be written as 

\begin{equation}
\log \mathcal{L}_{\mathrm{VI}} \sim \sum_{n=1}^{N} \log \mathcal{L}\left(\boldsymbol{Y}_{n}, \boldsymbol{\hat{Y}}_{n}\left(\boldsymbol{X}_{n}, \boldsymbol{\omega}\right)\right)-\lambda \sum_{i}\left\|\boldsymbol{\omega_{i}}\right\|^{2} ,
\label{eq:log_likelihood_vi_2}
\end{equation}

\noindent
where $\log \mathcal{L}\left(\boldsymbol{Y}_{n}, \boldsymbol{\hat{Y}}_{n}\left(\boldsymbol{X}_{n}, \boldsymbol{\omega}\right)\right)$ is the log-likelihood of the network predictions $\boldsymbol{\hat{Y}}_{n}\left(\boldsymbol{X}_{n}, \boldsymbol{\omega}\right)$ for training input $\boldsymbol{X}_n$ with true values $\boldsymbol{Y}_n$, $\lambda$ is the strength of the regularization term, and $\boldsymbol{\omega}_i$ are sampled from $q(\boldsymbol{\omega})$. 

For the multivariate Gaussian distribution predicted by \gampen{}, $\mathcal{N}(\boldsymbol{\mu}, \boldsymbol{\Sigma})$, we can write the log-likelihood of the network predictions (first-term on the right side in Equation \ref{eq:log_likelihood_vi_2}) as 

\begin{equation}
\log \mathcal{L} \propto  \sum_{n}  -\frac{1}{2}\left[\boldsymbol{Y}_{n}-\boldsymbol{\hat{\mu}}_{n}\right]^{\top} \boldsymbol{\hat{\Sigma}}_n^{-1}\left[\boldsymbol{Y}_{n}-\boldsymbol{\hat{\mu}}_{n}\right] -\frac{1}{2} \log [\operatorname{det}(\boldsymbol{\hat{\Sigma}}_n)] ,
\label{eq:effective_log_likelihood}
\end{equation}

\noindent
where $\boldsymbol{\hat{\mu}}_n$ and $\boldsymbol{\hat{\Sigma}}_n$ are the mean and covariance matrix of the multivariate Gaussian distribution predicted by \gampen{} for an image, $\boldsymbol{X}_n$. 

We train \gampen{} by minimizing the negative log-likelihood of the output parameters for the training set, which by combining Eqs. \ref{eq:log_likelihood_vi_2} and \ref{eq:effective_log_likelihood}, can be written as

\begin{equation}
- \log \mathcal{L}_{VI} \propto  \sum_{n} \frac{1}{2}\left[\boldsymbol{Y}_{n}-\boldsymbol{\hat{\mu}}_{n}\right]^{\top} \boldsymbol{\hat{\Sigma_n}}^{-1}\left[\boldsymbol{Y}_{n}-\boldsymbol{\hat{\mu}}_{n}\right] + \frac{1}{2} \log [\operatorname{det}(\boldsymbol{\hat{\Sigma_n}})] + \lambda \sum_{i}\left\|\boldsymbol{\omega_{i}}\right\|^{2} .
\label{eq:ap:final_loss_fn}
\end{equation}

\section{Additional Technical Details on \gampen{}}

In Table \ref{tab:network_layers}, we have outlined the various layers of the \gampen{} framework along with the important parameters of each layer and the corresponding activation functions.

\startlongtable 
    \begin{deluxetable*}{cccc}
    %\tablenum{3}
    \tablecaption{Structure of \gampen{} \label{tab:network_layers}}
    \tablecolumns{4}
    \tablehead{
    \colhead{\textbf{Order}} & \colhead{\textbf{Type of Layer}} & \colhead{\textbf{Layer Description}} & \colhead{\textbf{Activation Function}}
    }
    \startdata
        \hline
        \hline
        \multicolumn{4}{c}{Upstream Spatial Transformer Network} \\
        \hline 
        \hline
        1 & Input & Size: $239\times239$ & --  \\
        %\hline
        2 & Convolutional & No. of Filters: 64 $\vert$ Filter Size: 11 & ReLU\tablenotemark{a} \\
        %\hline   
        3 & Max-Pooling & Kernel Size: 3 $\vert$ Strides: 2 & -- \\
        %\hline
        4 & Convolutional & No. of Filters: 96 $\vert$ Filter Size: 9 & ReLU \\
        %\hline
        5 & Max-Pooling & Kernel Size: 3 $\vert$ Strides: 2 & -- \\
        %\hline
        6 & Fully Connected & No. of neurons: 32 & ReLU \\
        %\hline
        7 & Fully Connected & No. of neurons: 1 & Linear \\
        %\hline
        & & & \\
        \hline
        \hline
        \multicolumn{4}{c}{Downstream Morphological Estimation Network} \\
        \hline
        \hline
        1 & Input & Size: $239\times239$ & --  \\
        2 & Convolutional & No. of Filters: 64 $\vert$ Filter Size: 3 $\vert$ Strides: 1 & ReLU \\
        3 & Dropout & - & - \\
        4 & Convolutional & No. of Filters: 64 $\vert$ Filter Size: 3 $\vert$ Strides: 1 & ReLU \\
        5 & Max-Pooling & Kernel Size: 2 $\vert$ Strides: 2 & -- \\
        6 & Dropout & - & - \\
        7 & Convolutional & No. of Filters: 128 $\vert$ Filter Size: 3 $\vert$ Strides: 1 & ReLU \\
        8 & Dropout & - & - \\
        9 & Convolutional & No. of Filters: 128 $\vert$ Filter Size: 3 $\vert$ Strides: 1 & ReLU \\
        10 & Max-Pooling & Kernel Size: 2 $\vert$ Strides: 2 & -- \\
        11 & Dropout & - & - \\
        12 & Convolutional & No. of Filters: 256 $\vert$ Filter Size: 3 $\vert$ Strides: 1 & ReLU \\
        13 & Dropout & - & - \\
        14 & Convolutional & No. of Filters: 256 $\vert$ Filter Size: 3 $\vert$ Strides: 1 & ReLU \\
        15 & Dropout & - & - \\
        16 & Convolutional & No. of Filters: 256 $\vert$ Filter Size: 3 $\vert$ Strides: 1 & ReLU \\
        17 & Max-Pooling & Kernel Size: 2 $\vert$ Strides: 2 & -- \\
        18 & Dropout & - & - \\
        19 & Convolutional & No. of Filters: 512 $\vert$ Filter Size: 3 $\vert$ Strides: 1 & ReLU \\
        20 & Dropout & - & - \\
        21 & Convolutional & No. of Filters: 512 $\vert$ Filter Size: 3 $\vert$ Strides: 1 & ReLU \\
        22 & Dropout & - & - \\
        23 & Convolutional & No. of Filters: 512 $\vert$ Filter Size: 3 $\vert$ Strides: 1 & ReLU \\
        24 & Max-Pooling & Kernel Size: 2 $\vert$ Strides: 2 & -- \\
        25 & Dropout & - & - \\
        26 & Convolutional & No. of Filters: 512 $\vert$ Filter Size: 3 $\vert$ Strides: 1 & ReLU \\
        27 & Dropout & - & - \\
        28 & Convolutional & No. of Filters: 512 $\vert$ Filter Size: 3 $\vert$ Strides: 1 & ReLU \\
        29 & Dropout & - & - \\
        30 & Convolutional & No. of Filters: 512 $\vert$ Filter Size: 3 $\vert$ Strides: 1 & ReLU \\
        31 & Max-Pooling & Kernel Size: 2 $\vert$ Strides: 2 & -- \\
        32 & Fully Connected & No. of neurons: 4096 & ReLU \\
        33 & Dropout & - & - \\
        34 & Fully Connected & No. of neurons: 4096 & ReLU \\
        35 & Dropout & - & - \\
        36 & Fully Connected & No. of neurons: 9 & Linear \\
        \hline
    \enddata
    \tablenotetext{a}{Rectified Linear Unit}
    \tablecomments{The dropout rate of the various layers are set according to the callibration step described in \S \ref{sec:training}}
    \end{deluxetable*}

%\section{To Dos}
%\begin{itemize}
%    \item GET THE INTERPRETABILITY WORK DONE FOR THIS NETWORK IF YOU WANT TO INCLUDE THAT IN THIS PAPER
%    \item (Last pass)---Go over Daily Journal 2 to look over things you could add to this paper
%    \item (Last pass)---Go over the group meeting slides to look over things you could add to this paper
%\end{itemize}

\software{PyTorch \citep{pytorch},
          Ignite \citep{ignite},
          MLFlow \citep{mlflow},
          Numpy \citep{numpy},
          Astropy \citep{astropy_1,astropy_2},
          Pandas \citep{pandas},
          Scikit-learn \citep{scikitlearn},
          Matplotlib \citep{matplotlib},
          Corner \citep{corner},}

\bibliographystyle{aasjournal}
\bibliography{references}

\end{CJK*}
\end{document}

%% file: bangla_commands.tex
\font\bngxi=bang10 scaled 1100

\def\*#1*#2{o\null{#2}{#1}}

\def\d#1{\oalign{\smash{#1}\crcr\hidewidth{$\!$\rm.}\hidewidth}}

\def\sh#1{\setbox0=\hbox{#1}%
     \kern-.02em\copy0\kern-\wd0
     \kern.04em\copy0\kern-\wd0
     \kern-.02em\raise.0433em\box0 }